\documentclass[prl,aps,showpacs]{revtex4}
\usepackage{latexsym}
\usepackage{amssymb}
\usepackage{epsfig}
\usepackage{amsmath}
\usepackage{graphicx}

\begin{document}
\title{Bremsstrahlung from dense plasmas and the
  Landau-Pomeranchuk-Migdal effect}
\author{C. Fortmann, H. Reinholz, G. R\"opke, A. Wierling}
\affiliation{University of Rostock, Institut f\"{u}r Physik,
Universit\"atsplatz 3, D-18051 Rostock, Germany}
\date{\today}
\begin{abstract}
The suppression of the bremsstrahlung cross section due to multiple
scattering of the emitting electrons is an important effect in
dense media (Landau-Pomeranchuk-Migdal effect).
Here, we study the emission from a dense, fully-ionized and 
non-relativistic hydrogen plasma.   
Using the dielectric approach, we relate optical properties such as emission and absorption to
equilibrium force-force correlation functions, which allow for
a systematic perturbative treatment with the help of thermodynamic
Green functions. By considering self-energy and vertex corrections,
medium modifications such as multiple scattering of the emitting electrons
are taken into account. 
Results are presented for the absorption coefficient as a 
function of the frequency at various densities. It is shown that 
the modification of the inverse bremsstrahlung due to 
medium effects becomes more significant in the low frequency 
and high density region.
\end{abstract}


\maketitle

\section{1. INTRODUCTION}

In a fully ionized plasma, bremsstrahlung and inverse bremsstrahlung
are the only emission and absorption processes, respectively. For partially ionized plasmas these processes contribute to some extent
to the continuous emission/absorption spectrum. 
We introduce the emission coefficient
$j( \omega)$ as the
rate of radiated energy per unit volume, frequency and 
solid angle, and the absorption coefficient $ \alpha( \omega)
$ as the relative attenuation of 
the intensity of electromagnetic waves propagating in 
the medium per unit length. 
For a thermally equilibrated plasma, these quantities  
are linked by Kirchhoff's law \cite{griem}
\begin{equation} 
j( \omega)=L( \omega) \alpha( \omega)\;\;\;,
\end{equation} 
with the Planck distribution $L( \omega)=\hbar\omega ^3/\left[4\pi^3c^2
(\exp( \hbar\omega/k _{\mathrm{  B}}T)-1)\right]$.  
Thus, it is sufficient to study one or the other, i.e. $j(\omega)$ or
$\alpha(\omega)$ . Here, we choose the 
absorption coefficient $\alpha(\omega)$.

The absorption spectrum can be determined according to Quantum Electrodynamics (QED) 
\cite{itzykson}
from the interaction part of the QED Lagrangian
\begin{eqnarray}
\label{IntLagr}
{\mathcal L}^{}_{\rm int}(x)=
\sum_{c}  Z_c e j^{c}_{\mu}(x) A^{\mu}(x),  
\end{eqnarray}
$x$ being a four-dimensional space-time
variable.
This Lagrangian describes a minimal coupling between the 
particle current $j_{\mu}^c$ and the vector potential $A^{\mu}$.
The index $c$ denotes the species of the particles involved, carrying
the charge $Z_ce$.
Introducing a Hamiltonian $H$, the
transition rate 
between asymptotically free states of electrons
$|\,\boldsymbol{p}_\mathrm{in}\,\rangle, 
|\,\boldsymbol{p}_\mathrm{out}\,\rangle$ with the
energies $E^\mathrm{e}_\mathrm{in}, E^\mathrm{e}_\mathrm{out}$ respectively, follows from
Fermi's Golden rule as
\begin{eqnarray}
\label{GoldRule}
w_{\rm in,out}^{}
= \frac{2\pi}{\hbar} 
\Big|\langle \boldsymbol{p}_{\rm out} |H_\mathrm{int}| 
\boldsymbol{p}_{\rm in}^{}
\rangle\Big|_{}^{2} 
\delta(E_{\rm in}^{\rm e}+\hbar \omega-E_{\rm out}^{\rm e}) ~,
\end{eqnarray}
with $H_\mathrm{int}=-\int\mathrm{d} ^3x\,\mathcal L_\mathrm{int}(x)$.
Assuming single, uncorrelated scattering from different
ions only, the  
absorption coefficient is given by
\begin{eqnarray}
        \label{AbsorpCoeff}
        \alpha(\omega)
	=n_{\rm i}^{}\Omega_{0} \int 
        \frac{\Omega_{0} \mathrm{d} _{}^{3}p_{\rm in}^{}}{(2\pi)_{}^{3}}\,
        \frac{\Omega_{0} \mathrm{d} _{}^{3}p_{\rm out}^{}}{(2\pi)_{}^{3}}\;
        f(E^\mathrm{e}_{\rm in}) \frac{1}{c} w_{\rm in,out}^{} ~.
\end{eqnarray}
Here, $n_{\rm i}^{} \Omega_0$ is the total number of ions in the volume
$\Omega_{0}$ and 
$w_{\rm in,out}^{}$ is the transition probability that a photon of 
momentum $\boldsymbol{k}=\boldsymbol{p}_{\rm out} - \boldsymbol{p}_{\rm in}
$ and polarization $\lambda$ is absorbed by a single 
electron of momentum $\boldsymbol{p}_{\rm in}^{}$ in the Coulomb potential of 
an ion, leaving the process with momentum $\boldsymbol{p}_{\rm out}^{}$. 
The momentum
distribution function of the incoming electrons is denoted as
$f(E^\mathrm{e}_{\rm in})=\left[\exp\left( (\hbar^2 p^2_{\rm in}/2m_\mathrm{e} -\mu_\mathrm{e})/k_\mathrm{B}T\right)+1)\right]^{-1} $.  
The factor $1/c$ arises due to the current of incoming photons.

Evaluating the transition rate in Born approximation 
\begin{eqnarray}
        \langle \boldsymbol{p}_{\rm out} |H_\mathrm{int}| 
                \boldsymbol{p}_{\rm in} \rangle &  = &
        \frac{Z_ie^2 \hbar^2}{\epsilon_0 m_{\rm e} \Omega_0}
   \sqrt{\frac{e^2}{2 \hbar \epsilon_0 \Omega_0 \hbar \omega}}\,
   \frac{ \left( \boldsymbol{p}_{\rm out} -\boldsymbol{p}_{\rm in} \right)_z}{
   |{\boldsymbol p}_{\rm out}-{\boldsymbol p}_{\rm in}|^2}
\end{eqnarray} 
results in the well known Bethe-Heitler cross section 
\cite{beth:proc.roy.soc36,Heitler}. 

In the low frequency limit, the Bethe-Heitler cross section
 behaves roughly
like $1/ \omega$. Within QED, this infrared-divergence is 
discussed as a consequence of the neglect of vertex corrections \cite
{holstein}.  The infrared-divergent terms can be shown to cancel with
corresponding contributions to the form factor of the source particle
which is a manifestation of the Ward-Takahashi identities 
\cite{bjor:qfield}.

In the non-relativistic limit and for soft photons, 
the absorption coefficient for a hydrogen plasma ($Z_\mathrm{i}=1$)
is given by 
\begin{eqnarray}
  \label{eq:alpha_born}
  \alpha^{\rm B}(\omega) & = & 
  \frac{C}{\hbar \omega^3}\,n_{\rm e}\,n_{\rm i}\,
  \sinh\left(\frac{\hbar \omega}{2 k_{\rm B} T} \right)\,
  K_0\left( \frac{\hbar \omega}{2 k_{\rm B} T} \right)~,
\end{eqnarray}
where $C^{-1}=3 \sqrt{2} \epsilon_0^3 \pi^{3/2}
m_{\rm e}^{3/2}c \left( k_{\rm B} T \right)^{1/2}/e^6$ and 
$K_0(x) =\int_0^\infty \cos(x \sinh t) \mathrm{d} t$ is the modified 
Bessel function.
Here, the electron density $n_\mathrm{e}$ has been introduced, which is equal to the ion density $n_\mathrm{e}=n_\mathrm{i}$ 
in charge neutral systems.

This Born approximation can be improved by taking Coulomb wave
functions for the  initial and final states. In this case, 
an analytic result for the absorption coefficient 
was given by Sommerfeld \cite{S}.
Due to the occurrence of 
hyper-geometric functions, simpler approximations such as 
the Born-Elwert approximation \cite{E} have 
been developed. In the classical limit, the Sommerfeld expression
reduces to a result obtained by Kramers earlier \cite{K}.

In a dense plasma, 
the influence of the collective
behavior of the system and the modification of single-particle 
properties of the emitting and absorbing particles as well as the
bremsstrahlung photons are important.
Retaining 
the single-scattering picture of Eq.~(\ref{AbsorpCoeff}), medium
effects can be taken into account 
e.g. by a modification of the potential. 
Instead of a Coulomb potential, a static or dynamically screened 
potential should be  used \cite{B1}. A quantum-statistical approach
based on a systematic perturbative treatment of the force-force
correlation function has been developed in Ref.~\cite{rein:pre00}, 
for an application to inverse bremsstrahlung see Ref.~\cite{wier:phpl01}.

How does multiple scattering of the emitting electron change 
the cross section? 
This question was
first treated by Landau and Pomeranchuk \cite{land:dokl53} in a
semi-classical way and soon afterwards by Migdal \cite{migd:physrev56}
using quantum statistical methods. They showed
that the account of successive collisions leads to
a suppression of the bremsstrahlung cross section compared to the Bethe-Heitler 
result at photon energies low against the energy of
the scattering electron. 
Both Bethe-Heitler and Landau-Pomeranchuk-Migdal (LPM) theory 
give the same result in the limiting
cases of high photon energies and/or low
densities, i.e., in those cases where the Born approximation is applicable. 
Migdal's result has been rederived more recently 
using different methods, e.g., path-integral calculations 
\cite{zakh96,zakh98} and quantum kinetic equations \cite{kosh:jphysa02}.
A comprehensive overview of theoretical approaches is given 
in \cite{klei:revmodphys99}. There, it is also pointed out that the LPM effect
might play an important r\^ole for the emission/absorption spectrum 
of a plasma even in the non-relativistic regime due to the large 
number of free charge carriers. Other aspects of the emission of radiation 
are also under consideration such as a coherent
state description of photon emission
\cite{wang:jphys01,czac:phle03}. 

Due to the large energies required to observe notable effects, 
it took quite a long time until experiments could
unambiguously approve LPM theory. 
Experimental investigations have been performed since the late 1950's
using high energy
electrons (some MeV to GeV) from cosmic rays
\cite{fowl:phil.mag59,varf:jetp60,lohr:pr61,kasa:prd85,stra91} 
and accelerators \cite{varf:jetp75}.
These early experiments 
suffered from poor 
statistics and were unable to confirm LPM theory. 
More recent experiments at SLAC \cite{anto:prd97} and CERN 
\cite{bak:nucl88,hans:prl03,hans:prd04} have indeed shown the LPM effect.

Migdal's theory is not completely microscopic but relies on the
definition of a macroscopic parameter, namely the coherence length
$l$, initially introduced by Ter-Mikaelyan \cite{klei:revmodphys99}
and first applied to the theory of bremsstrahlung by Landau and
Pomeranchuk \cite{land:dokl53}. The coherence length gives the scale
of photon energy, below which the suppression of bremsstrahlung
becomes important.  It is basically determined by the density of the
medium.
Knoll and Voskresensky \cite{knol:annals96} were the first to
treat the question of the LPM effect in the context of a many particle system, where
every constituent has to be regarded as an emitter of bremsstrahlung.
They were able to show a suppression of the emission/absorption spectrum at
low frequencies by using medium modified single particle
propagators. The particles are assigned a finite lifetime $\tau$,
given by the width of their spectral function
$\Gamma=\hbar/\tau$. However, this quantity is not calculated from a
microscopic approach, but simply set as a parameter. It is related to
the aforementioned coherence length by the simple relation $\tau\simeq
l/c$ \cite{voskr_pc03}.

In this work, we present a completely
microscopic calculation of the absorption spectrum, where 
the spectral function of the electron is obtained in a self
consistent manner.  Thus, we present for the first time a fully microscopic
treatment of the LPM effect for a many-particle system.
The absorption coefficient is defined in linear response theory and is expressed through 
thermodynamical correlation functions using a diagram technique 
equivalent to the well known Feynman diagrams \cite{krae}. Medium 
effects are accounted for systematically in terms of 
self-energy and vertex corrections.
We show that the Bethe-Heitler bremsstrahlung spectrum follows 
from this approach in lowest order perturbation theory using 
free particle propagators.
The account of coherent successive scattering, as in LPM theory, 
can be achieved by a partial summation of 
self-energy diagrams. This procedure leads to a finite width of the
single-particle spectral function which 
reflects the modification of the
energy-momentum dispersion relation due to the medium.
It is found that the absorption coefficient calculated on the basis of these
medium modified propagators is significantly altered 
in the low frequency range in comparison to the Born 
approximation. In the high frequency limit as well as in the low density
limit, the Born approximation is reproduced. 
Thus, the main features of LPM theory can be rederived within our microscopic approach.

\section{2. LINEAR RESPONSE THEORY}
We consider the interaction of soft photons with a non-relativistic,
homogeneous 
plasma. 
A key quantity to describe the propagation of electro-magnetic 
waves in a medium
is the dielectric tensor $\epsilon_{ij}({\boldsymbol k},\omega)$ \cite{jack}.
In the isotropic case, the tensor can be decomposed into a transverse
$\epsilon_{\rm t}(k,\omega)$ 
and a longitudinal $\epsilon_\mathrm{l}(k,\omega)$
part with respect to the wave vector $\boldsymbol{k}$. Here and in the
following, we take $\boldsymbol k$ to point along the $z$-axis,
$\boldsymbol k = k \boldsymbol e_z$.
In the long-wavelength limit $\boldsymbol k \to 0$, the longitudinal and the
transverse part coincide
$\epsilon(\omega) = \lim_{k \to 0} \epsilon_\mathrm{l}(k,\omega)=
\lim_{k \to 0} \epsilon_\mathrm{t}(k,\omega)$.
The absorption coefficient can be
obtained from the dielectric function according to
\begin{eqnarray}
  \label{eq:alpha_eps}
  \alpha(\omega) & = & \frac{\omega}{c} \frac{{\rm Im}\,
    \epsilon(\omega)}{n(\omega)}   
   \;\;\;,
\end{eqnarray}
where the index of refraction $n(\omega)$ is also linked to the dielectric
function by
\begin{eqnarray}
  \label{eq:n_eps}
  n(\omega) & = & 
  \frac{1}{\sqrt{2}}\left( {\rm Re }\, \epsilon(\omega) + \left| \epsilon(\omega) \right|
  \right)^{1/2}\;\;\;.
\end{eqnarray}
The relation between $\epsilon_\mathrm{l}(k,\omega)$ and the 
longitudinal
response function $\chi^{}_\mathrm{l}(k,\omega)$
\begin{eqnarray}
  \label{eq:eps_chi}
  \epsilon_\mathrm{l}(k,\omega )=\frac{1}{1-\frac{e^2}{\epsilon_0k^2}\chi^{}_\mathrm{l}(k,\omega )}~,
\end{eqnarray}
allows for a microscopic approach to the dielectric function. 
Within linear response theory, the Kubo formula relates the response 
function to the current-current correlation function \cite{maha}
\begin{eqnarray}
  \label{eq:Kubo}
  \chi^{}_{\rm l}({\boldsymbol k},\omega) & = &  i \beta \Omega_0 
  \frac{k^2}{\omega}\,
  \langle J_k^z; J_k^z \rangle_{\omega+i  \eta} \;\;\;,
\end{eqnarray}
where the correlation functions for two observables $A,B$ 
are defined according to 
\begin{eqnarray}
        (A;B) &=& {1 \over \beta} \int\limits_0^\beta \mathrm{d} \tau\, 
        {\rm Tr}[A(-i  \hbar \tau) B^\dagger \rho_0]
        \;\;\;, \nonumber\\
        \langle A;B \rangle_{\omega +i \eta}  &=& 
        \int\limits_0^\infty \mathrm{d} 
        t\,\mathrm{e}^{i (\omega +i \eta)t}\, 
        (A(t);B)~.
\end{eqnarray}
The time dependency of the operators is taken in the Heisenberg picture.
The current density operator is given as
\begin{equation}
        {\boldsymbol J}_k 
        = {1 \over \Omega_0} \sum_{c,\boldsymbol{p}} {e_c \over m_c} 
        \hbar \boldsymbol {p} \, a^\dagger_{c, \boldsymbol{p} - 
        \boldsymbol{k}/2} 
        a^{}_{c,\boldsymbol{p} + \boldsymbol{k}/2}\;\;\;.
\end{equation}
$a^\dagger_{c,p}$ and $a_{c,p}$ are creation and annihilation 
operators for momentum states, respectively. 
$c$ labels the species and further quantum numbers such as 
spin, $\Omega_0$ is a normalization volume,
$\rho_0$ is the equilibrium statistical operator,
$\beta=1/(k_{\rm B} T)$ is the inverse temperature. Note, that in 
Eq.(\ref{eq:Kubo}), a small but finite imaginary part $\eta$
has been added. In the final results, the limit $\eta \to 0$ is taken.\\
The inverse response function can also be expressed as \cite{rein:pre00} 
\begin{eqnarray}
        \chi^{-1}_{\rm l}(\boldsymbol{k}, \omega) =  {i 
        \over \beta \Omega_0} {\omega \over
        k^2} {1 \over ( J^z_k ; J^z_k)^2 } \left[-i 
        \omega ( J^z_k ; J^z_k) +
        \langle \dot J^z_k ; \dot J^z_k \rangle_{\omega + i 
        \eta} - {\langle
        \dot J^z_k ; J^z_k \rangle_{\omega + i  \eta} 
        \langle J^z_k ; \dot
        J^z_k \rangle_{\omega + i  \eta} 
        \over \langle J^z_k ; J^z_k
        \rangle_{\omega + i  \eta}} \right] \;\;\;.
\end{eqnarray}
This transformation of the current-current correlation function into a
force-force correlation 
        function $\langle \dot J^z_k ; \dot J^z_k \rangle_{\omega + i 
        \eta}$ with 
	\begin{equation}
		\boldsymbol{\dot{J}}=
		\frac{i}{\hbar}\left[H,\boldsymbol{J}\right]~,
	\end{equation} 
	which has the meaning of a force as the time derivative of momentum,
	is more suited for a perturbative
treatment \cite{rein:pre00}. Also, it is convenient to introduce a 
generalized collision frequency $\nu(\omega)$
in analogy to the Drude relation \cite{rein:pre00}
\begin{eqnarray}
        \label{eq:gen_drude}
        \epsilon(\omega) & = & 1\,-\, \frac{\omega_{\rm pl}^2}{
        \omega\,\left( \omega+ i \nu\left(\omega\right) \right)
        } \;\;\;,
\end{eqnarray}
where $\omega_{\rm pl}^2=\sum_c\, n_ce_c^2/(\epsilon_0m_c)$ 
is the squared plasma frequency. \\
By comparison with Eq.~(\ref{eq:eps_chi}), we establish an expression 
for the collision frequency in terms of correlation functions
\begin{equation}
        \label{Def:nu}
        \nu(\omega) = \frac{\beta \Omega_0}{\epsilon_0\omega_{\rm pl}^
        {2}}\,
        \lim_{k \to 0}\,\left[
        \big\langle \,\,\dot{\!\!J}_{k}^{z} , \,\,\dot{\!\!J}_{k}^{z} 
        \big\rangle_{\omega+i \eta}\,-\,
        \frac{\langle \dot J^z_k ; J^z_k \rangle_{\omega + i 
        \eta}\,
        \langle J^z_k ; \dot J^z_k \rangle_{\omega + i  \eta}}
        { \langle J^z_k ; J^z_k \rangle_{\omega + i  \eta}} 
        \right]
        \;\;\;,
\end{equation}
taking into account that $( J^z_k ; J^z_k)= 
\epsilon_0 \omega_{\rm pl}^2/\beta \Omega_0$ is an exactly known 
quantity. Further details can be found in Ref.~\cite{rein:pre00}.
Making use of Eq.~(\ref{eq:alpha_eps}), 
the absorption coefficient can be expressed as
\begin{equation}
        \alpha(\omega)= \frac{\omega_{\rm pl}^{2}}{c}
        \frac{{\rm Re}\;\nu(\omega)}{(\omega_{}^{2}-2 \omega\, 
        {\rm Im}\,\nu(\omega) + |\nu(\omega)|^2)
        n(\omega)}\;\;\;.
\end{equation}

In the high frequency limit $\omega \gg \omega_{\rm pl}$,
the index of refraction is unity and the collision frequency is small
compared to the frequency $\omega$. Then, we can consider the
approximation
\begin{eqnarray}
        \label{Def:alpha}
        \alpha(\omega) & = & \frac{\omega_{\rm pl}^2}{c\,\omega^2}\,
        \mbox{Re}\, \nu(\omega) 
        \; = \;  \frac{\beta \Omega_0}{c \epsilon_0 \omega^2}\,
        {\rm Re}\, \big\langle \,\,\dot{\!\!J}_{0}^{z} , \,\,\dot{\!\!J}_
        {0}^{z} 
        \big\rangle_{\omega+i \eta}\;\;\;,
\end{eqnarray}
where the collision frequency is given in the form of a force-force
correlation function, cf.~Ref.~\cite{rein:pre00}. Thus, the absorption
coefficient is directly proportional to the 
real part of the force-force correlation function, which itself
can be determined using perturbation theory. 

As well known, the deviation of the diffraction index $n(\omega )$ from unity
at frequencies near the plasma frequency $\omega _\mathrm{pl}$
is responsible for the so-called dielectric
suppression of the bremsstrahlung spectrum. We refer to the pioneering work of
Ter-Mikaelyan \cite{ter}. In our approach, making use of Eqs.
(\ref{eq:n_eps}), (\ref{eq:gen_drude}), and  
(\ref{Def:nu}), the index of refraction can be determined from the
force-force correlation function. However, due to the choice of the
frequency range in consideration ($\omega \gg \omega _\mathrm{pl}$),
this effect will not be considered in the
present work. We will focus only
on the medium effects obtained directly from the evaluation of the
force-force correlation function.
\section{3. GREEN FUNCTION APPROACH AND DENSITY EFFECTS}
        
\begin{figure}
	\includegraphics[width=.8\textwidth,clip]{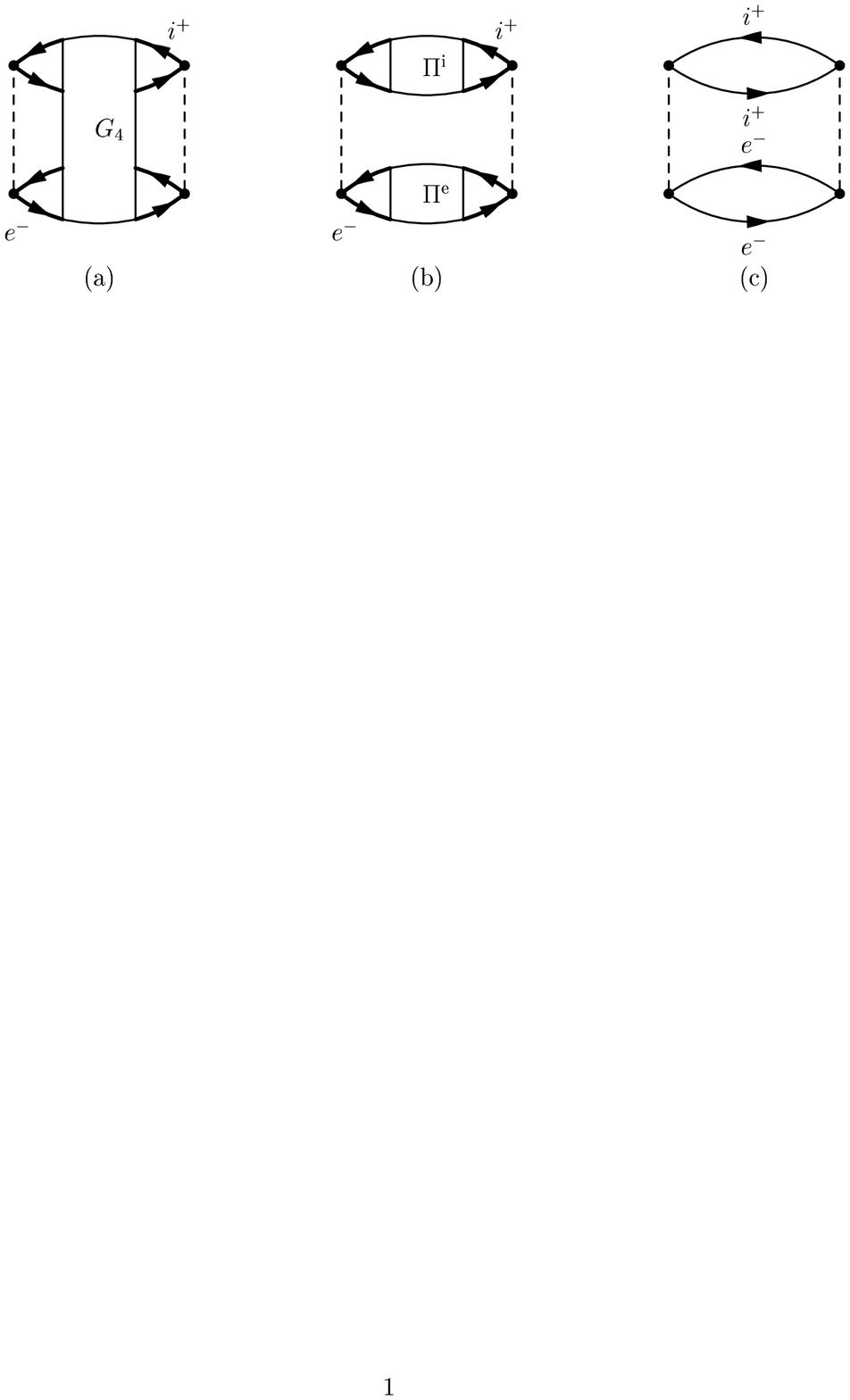}
	\caption{Diagrammatic representation of $G_{\dot J \dot J}
	(\omega_{\mu})$. (a) full account of all
	medium effects by a four particle Green function,
	\\(b) factorization into two polarization bubbles,
	(c) Born approximation.
	\label{fig1}}
\end{figure}

A convenient starting point for a perturbative treatment of the
force-force correlation function is the representation in terms of
a Green function $G_{\dot J \dot J}$ in the limit $k \to 0$ \cite{zuba2}
\begin{eqnarray}
        \label{eq:corr_green}
        \langle \dot J_0^z, \dot J_0^z \rangle_{\omega+i  \eta}
        & = & 
        \frac{i }{\beta} \int_{-\infty}^{\infty}\!\frac{\mathrm{d} 
        \bar \omega}{\pi}\,
        \frac{1}{\omega+i  \eta -\bar \omega} 
        \,\frac{1}{\bar \omega}\,
        {\rm Im}\, G_{\dot J \dot J}(\bar \omega+i\eta)~.
\end{eqnarray}
By exploiting Dirac's identity 
\begin{eqnarray}
\label{eq:dirac}
\lim_{\eta\to 0}\frac{1}{x \pm i  \eta} & = & {\cal P}\frac{1}{x} \mp 
i \pi \delta(x)~,
\end{eqnarray}
the real part of the force-force correlation function is evaluated to
\begin{equation}
        {\rm Re}\,\langle \dot J_0^z, \dot J_0^z \rangle_{\omega+i  \eta}=
        \frac{1}{\beta\omega }{\rm Im}\,G_{\dot J\dot J}(\omega+i\eta )~.
        \label{force-force-correl-green}
\end{equation}
Thus, the absorption coefficient reads in the high-frequency limit
\begin{eqnarray}
\label{eq:alpha_green}
  \alpha(\omega) & = & 
  \frac{ \pi \Omega_0}{c \epsilon_0 \omega^3} \,{\rm Im}
  \,G_{\dot J \dot J}(\omega+ i  \eta)~.
\end{eqnarray}
The time derivative of the electron current density operator is calculated as
\begin{equation}
        \label{eqn:time_deriv_current_hamiltonian}
        \dot J^z_{0,{\rm e}}  = \frac{i }{\hbar}\big[H,J^z_{0,{\rm
            e}}\big]
        = 
	\frac{i  e}{m_\mathrm{e} \Omega_0}\,\sum_{\boldsymbol{pkq}} 
        v_q^{\rm ei} 
        q_z 
	a^{\dagger}_{\mathrm{e}, \boldsymbol{p}} \,a^{\dagger}_{\mathrm i, \boldsymbol{k}}\, 
	a_{\mathrm i, \boldsymbol{k}-\boldsymbol{q}}  \,a_{\mathrm{e}, \boldsymbol{p}+\boldsymbol{q}}        
\end{equation}
with the Hamiltonian
\begin{eqnarray}
	H&=& \sum_{c,\boldsymbol{k}} 
E^c_k
a^\dagger_{c,\boldsymbol{k}}a^{}_{c,\boldsymbol{k}}
+\frac{1}{2}\sum_{c,d\atop \boldsymbol{kpq}}v^{cd}_qa^\dagger_{c,\boldsymbol{k}+\boldsymbol{q}}
a^{\dagger}_{d,\boldsymbol{p}-\boldsymbol{q}}a^{}_{d,\boldsymbol{p}}a^{}_{c,\boldsymbol{k}}~
        \label{eqn:hamiltonian_a-op}
\end{eqnarray}
and $E_k^c= \hbar ^2 k^2/2 m_c$. The spin is not given explicitly but
is included into the free-particle quantum number $c$. Due to conservation  
of total momentum of electrons, only electron-ion collisions
contribute to Eq.(\ref{eqn:time_deriv_current_hamiltonian}).
With Eq.(\ref{eqn:time_deriv_current_hamiltonian}), 
we identify the Green function as a four particle Green function.
Its diagrammatic representation is shown in FIG.~\ref{fig1}, l.h.s.
Here, $G_4$ denotes a four-particle Green function that contains all
interactions between the considered particles \cite{wier:phpl01}.
We perform a sequence of approximations by selecting certain
diagrams contributing to $G_4$.
Considering the electron-ion interaction determining the force $\dot
J^z_{0,{\rm e}}$ only in lowest order, i.e. in Born approximation, but
the full correlations within the electron and ion subsystem,
respectively, we are led to the middle of FIG. 1 (b) showing only
those diagrams which can be factorized into two polarization bubbles.
$\Pi_\mathrm{e}$ denotes the electronic polarization function,
$\Pi_\mathrm{i}$ is the corresponding ionic quantity.
In this way, we keep the description of the 
single scattering event on the level of the Born
approximation. However, higher order interactions between the full
electron and ion subsystem such as a ladder approximation (t-matrix)
are ignored, 
cf. also Ref.~\cite{rein:pre00}. The t-matrix corrections and in
particular the reproduction of the Sommerfeld result \cite{S}
have been studied in Ref.~\cite{wier:phpl01}.
\subsection{A. Born approximation}
As a simple example and prerequisite for further improvements, we consider the Born approximation.  
In this case, $G_4$ is a product of four single particle propagators
and 
all single particle propagators are replaced by free
propagators. 
We obtain for the Green function the diagram given in 
FIG. \ref{fig1}, r.h.s.
Details of the calculation are discussed in App. A.
For a
Maxwellian plasma ($f(E^\mathrm{e}_p)=n_\mathrm{e}\Lambda_\mathrm{e}^3/2\,
\exp(-\beta E^\mathrm{e}_p)$),
we obtain 
\begin{eqnarray}
        \label{eqn:born_110}
        {\rm Im}\,G^\mathrm{Born}_{\dot J\dot J}(\omega )&=& 
        \frac{n_{\rm i} n_{\rm e} \Lambda_e^3e^6(1 - e^{-\beta \hbar
            \omega})}{24 \pi^3\epsilon_0^2 \hbar^4\omega ^2 } 
            \int\!\mathrm{d} E^\mathrm{e}_p\, e^{-\beta E^\mathrm{e}_p}
        \times\nonumber\\&&\times
	\left(-2\frac{\sqrt{E^\mathrm{e}_p}\ \sqrt{\hbar\omega + E^\mathrm{e}_p}\,\hbar^2\kappa^2/m_\mathrm{e} }{(\hbar\omega +\hbar^2\kappa^2/2m_\mathrm{e})^2+2E^\mathrm{e}_p\,\hbar^2\kappa^2/m_\mathrm{e}}
        +\frac{1}{2}\ln\left|
        \frac{\left(\sqrt{E^\mathrm{e}_ p+\hbar\omega }+\sqrt{E^\mathrm{e}_ p} \right)^2 + \hbar^2\kappa^2/2m_\mathrm{e}}{\left(\sqrt{E^\mathrm{e}_ p+\hbar\omega }-\sqrt{E^\mathrm{e}_ p} \right)^2 + \hbar^2\kappa^2/2m_\mathrm{e}}
        \right| \right).
\end{eqnarray}
Note, that the spin-degeneracy factor $1/2$ for fermions is compensated by
a factor $2$ from the summation over spin variables 
in the calculation of the correlation functions.
$\Lambda_\mathrm{e}=(2 \pi \hbar^2/m_\mathrm{e}k_\mathrm{B}T)^{1/2}$ is
the thermal de-Broglie wavelength.
$\kappa$ occurs due to the use of a statically 
screened potential of Debye-H\"uckel type
\begin{equation}
        \label{eqn:debye-pot}
        v_q^{\rm ei}=-\frac{Z_{\rm i}e^2}{\epsilon_0 \Omega_0
          (q^2+\kappa^2) }~, 
\end{equation}
with  $\kappa^2=\sum_c n_cZ^2_ce^2/(
\epsilon_0k_\mathrm{B}T)$,
to ensure
convergence at $\omega=0$. For $\omega \neq 0$, we can consider the
Coulomb limit $\kappa=0$. Performing the integration with the help of
\begin{equation}
        \int\limits_0^\infty {\mathrm{d} x \over x} \mathrm e^
        {-(a/x -bx)^2} = {\rm e}^{2ab} K_0(2ab)\;\;\;,
\end{equation}
we arrive at
Eq.~(\ref{eq:alpha_born}). Note, that this result is sometimes also 
written as \cite{Hutch}
\begin{equation}
        \alpha^{\rm Born}(\omega)= {n_{\rm i} n_{\rm e} 
        \Lambda_\mathrm{e}^3 e^6 \over 24 
        \hbar^4 \omega^3
        \epsilon_0^3 c \pi^3}(1-\mathrm e^{-\beta \hbar \omega}) 
        \int\limits_{0}^{\infty} \mathrm{d} E^\mathrm{e}_p\,
        \mathrm{e}^{-\beta E^\mathrm{e}_p} \ln 
        \left({\sqrt{E^\mathrm{e}_p+\hbar \omega}+\sqrt{E^\mathrm{e}_p} \over
        \sqrt{E^\mathrm{e}_p+\hbar \omega} - \sqrt{E^\mathrm{e}_p}} \right)~.
        \label{eqn:alpha_born_result}
\end{equation} 
It is instructive to study the different terms in 
Eq.(\ref{eqn:alpha_born_result}) in more detail: The integrand contains the
distribution function (Maxwell distribution). Furthermore, a
logarithm that depends on both electron and photon energy appears. Taken with appropriate prefactors, 
this logarithm is equal to the differential cross section for
inverse bremsstrahlung in the non-relativistic limit (Bethe-Heitler formula) \cite{jack}. 
Thus, we have a quite reasonable relation between the absorption
coefficient as a macroscopic property and the underlying microscopic 
process, namely 
inverse bremsstrahlung. The absorption spectrum is obtained through
integration 
of the cross section of the microscopic process weighted with the 
distribution function of the absorbing particles.\\
The absorption coefficient in 
Born approximation suffers the same logarithmic divergence 
in the limit $\omega \to 0$ (\emph{infrared divergence}) as the
non-relativistic limit of the Bethe-Heitler formula. \\
We will now show how the Born approximation can
be improved. As already mentioned above, improvements based on a more sophisticated
description of the single  scattering process via a t-matrix approach have been
obtained recently \cite{wier:phpl01}. Effects due to dynamical screening have also been considered. 
Both effects remove the
infrared divergence mentioned before.
In this work,  we want
to include medium effects such as the successive scattering of the
absorbing particles on ions during the absorption of the photon, 
a process becoming more important with
increasing density. This is also the basic idea of the LPM effect. 

The correction in lowest order to Born which takes account of medium effects in the propagator
can be achieved by performing either one self-energy insertion or one 
vertex correction in the sense of Ward-Takahashi identities.
In diagrammatic terms, this means
\begin{equation}
        \label{eqn:ward}
	\parbox{\textwidth}{
		\includegraphics[width=\textwidth,clip]{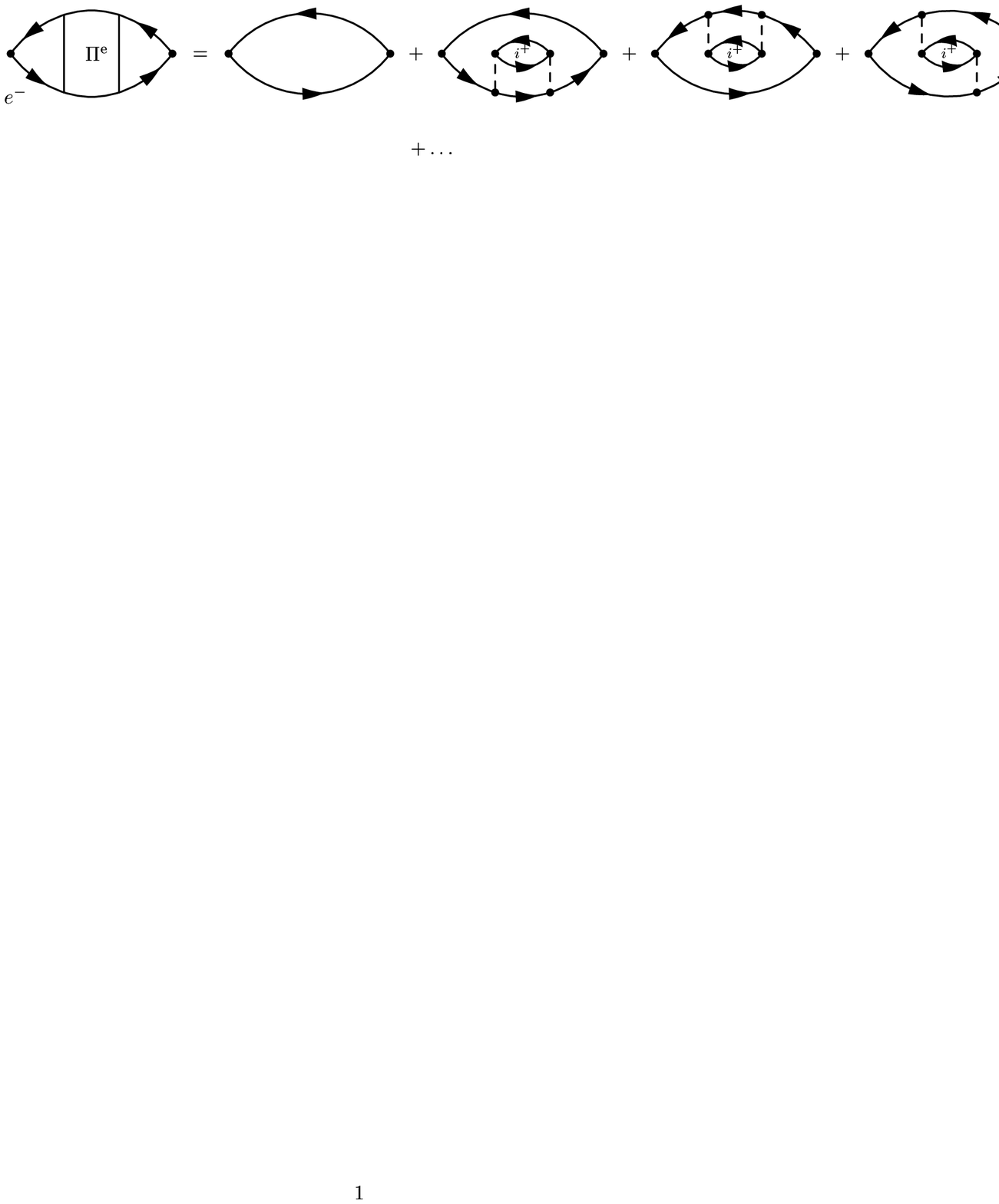}
		}
\end{equation}\\
We use $i^+$ to indicate loops
with ionic propagators.  
The first loop on the r.h.s.  yields the Born result as shown 
previously, in the
following  we have the self-energy and 
the vertex correction, respectively.\\
However, one does not obtain a finite result from this
ansatz in the case of the self-energy correction. Instead, 
a partial summation of all self-energy terms leading to a spectral 
function is necessary. 
The results are presented in subsection 3.C. In constrast, the last diagram of Eq.~(\ref{eqn:ward}) describing the
vertex correction gives a finite contribution as shown in subsection 3.D. 
A full self-consistent treatment of the vertex, i.e. solving the corresponding Bethe-Salpeter equation \cite{krae},
has not yet been performed.

\subsection{B. Spectral function}
We now discuss the influence of multiple scattering of the
source particles (electrons) on ions.
This can be accounted for by using dressed propagators in
the calculation of the force-force correlation function, i.e.
by replacing the free electron Green function by the expression
\begin{equation}
        G_{\mathrm e}(\boldsymbol p, z_\nu)=
        \int\limits_{-\infty}^\infty
        \frac{\mathrm{d} \hbar\omega }{2\pi}\frac{A_{\rm e}
        (\boldsymbol p,\hbar\omega) }
        {z_\nu-\hbar\omega }~,
        \label{eqn:spectral_repres}
\end{equation}
where $A_\mathrm{e}(\boldsymbol p,\hbar\omega)$ is the electronic 
spectral function. According to
Dyson's equation \cite{krae}, the Green function
can be represented by a complex electron 
self-energy $\Sigma_\mathrm{e}(\boldsymbol p, z_\nu)$,
\begin{eqnarray*}
	G_{\mathrm e}(\boldsymbol{p},z_\nu)&=&
	\parbox{.6\textwidth}{
	\includegraphics[width=.6\textwidth,clip]{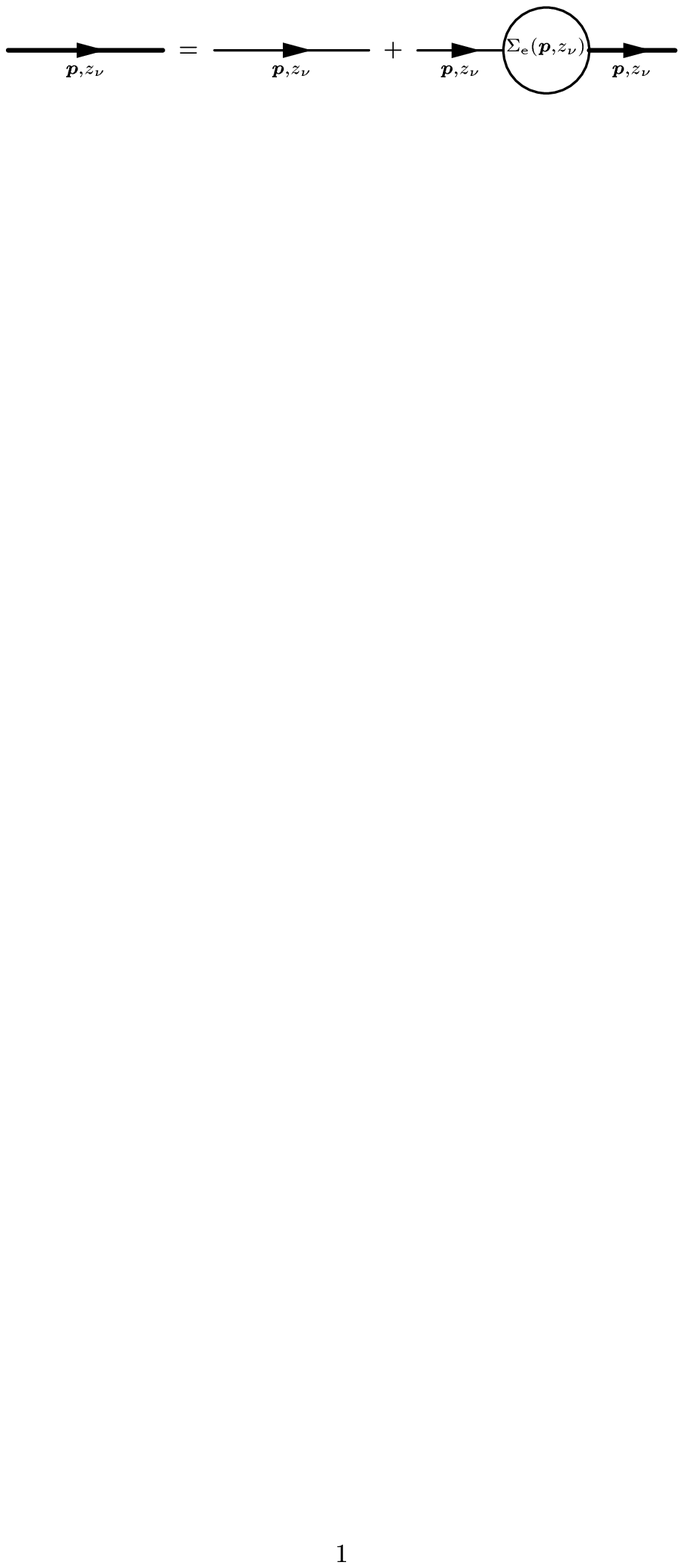} }
	\\ \\ &=& 
	\frac{1}{z_\nu-E^{\mathrm e}_p-\Sigma_\mathrm{e}(\boldsymbol p,z_\nu)}~.
\end{eqnarray*}
We perform the analytic continuation of the discrete Matsubara frequencies into
the upper half of the complex energy plane via $z_\nu\to\hbar\omega +i\eta$ 
and decompose the self-energy into the real and
imaginary part, $\Sigma_\mathrm{e}(\boldsymbol p,\hbar\omega+i\eta )= 
        \Delta_\mathrm{e}(\boldsymbol p,\hbar\omega)-
        i \Gamma_\mathrm{e}(\boldsymbol p,\hbar\omega)/2$. Then, the
        spectral function is related to the self-energy:
\begin{equation}
        A_\mathrm{e}(\boldsymbol p,\hbar
        \omega)=\frac{\Gamma_\mathrm{e}(\boldsymbol
          p,\hbar\omega)} 
        {(\hbar\omega -E^\mathrm{e}_p
        -\Delta_\mathrm{e}(\boldsymbol p,\hbar\omega))^2
        +\Gamma^2_\mathrm{e}(\boldsymbol p,\hbar\omega)/4}
        \label{eqn:def_sf}~.
\end{equation}
For the self-energy, we describe 
the scattering of the electron on an ion by a statically screened ion potential, 
cf. Eq.(\ref{eqn:debye-pot}).\\
The Hartree term vanishes due to charge neutrality.
The Fock term of the electron self-energy is not relevant, since we assume a non degenerate system.
The self-consistent first loop correction to the
Hartree-Fock self-energy due to the electron-ion interaction is given
by the diagram\\[.5cm]
\begin{equation}
        \Sigma_\mathrm{e}(\boldsymbol p,z_\nu)\ 
        =\  
	\parbox{.13\textwidth}{
	\includegraphics[width=.13\textwidth,clip]{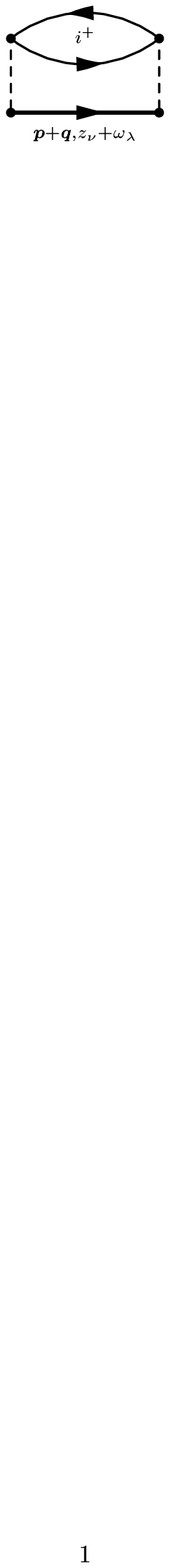}}
~.
       \label{eqn:sel_diagram}
       \end{equation}
       \\\\
This approximation can be improved by taking into account 
a partial summation of further loops, leading to the GW approximation
for the self-energy \cite{maha}.\\
       After analytic continuation of the Matsubara Green function
       we have
       \begin{equation}
       \Sigma_{\rm e}(\boldsymbol p,\hbar\omega+i\eta)
        = n_{\rm i} \int {\mathrm{d} ^3 q \over (2 \pi)^3} 
        \left|v_q^{\rm ei}\right|^2 
        {1 \over \hbar \omega+i \eta - 
        E_{ \boldsymbol{p+ q}}^{\rm e}-
	\Sigma_{\rm e}(\boldsymbol{p+q},\hbar \omega+i\eta)}~.
        \label{eqn:se_result}
	\end{equation}
Eq.~(\ref{eqn:se_result}) can be solved numerically by iteration 
starting from a suitable initialization.

\begin{figure}
	\includegraphics[width=.8\textwidth,clip,angle=-90]{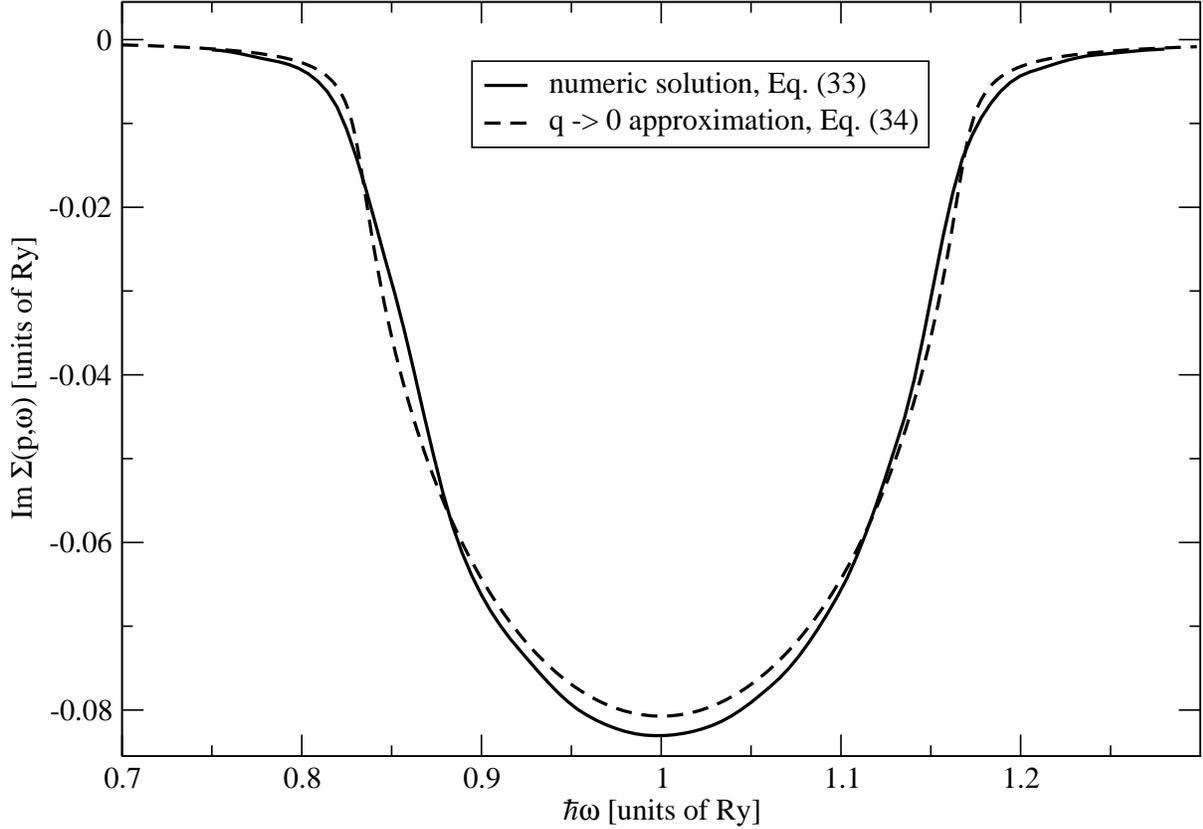}
	\caption{Solution of Eq.~(\ref{eqn:se_result}) for the 
	imaginary part of the self-energy. Both, the full solution and
	the approximation with $q=0$ coincide to a 
	large extent. The wave vector $p$ is taken as
	$p=1\,a_B^{-1}$. The parameters are $n_\mathrm{e}=10^{-6}\,a_\mathrm{B}^{-3}$ and $ k_\mathrm{B}T=2\ \mathrm{Ry}=27.2\ \mathrm{eV}$.
	\label{fig2}}
\end{figure}

\begin{figure}
	\includegraphics[width=.8\textwidth,clip,angle=-90]{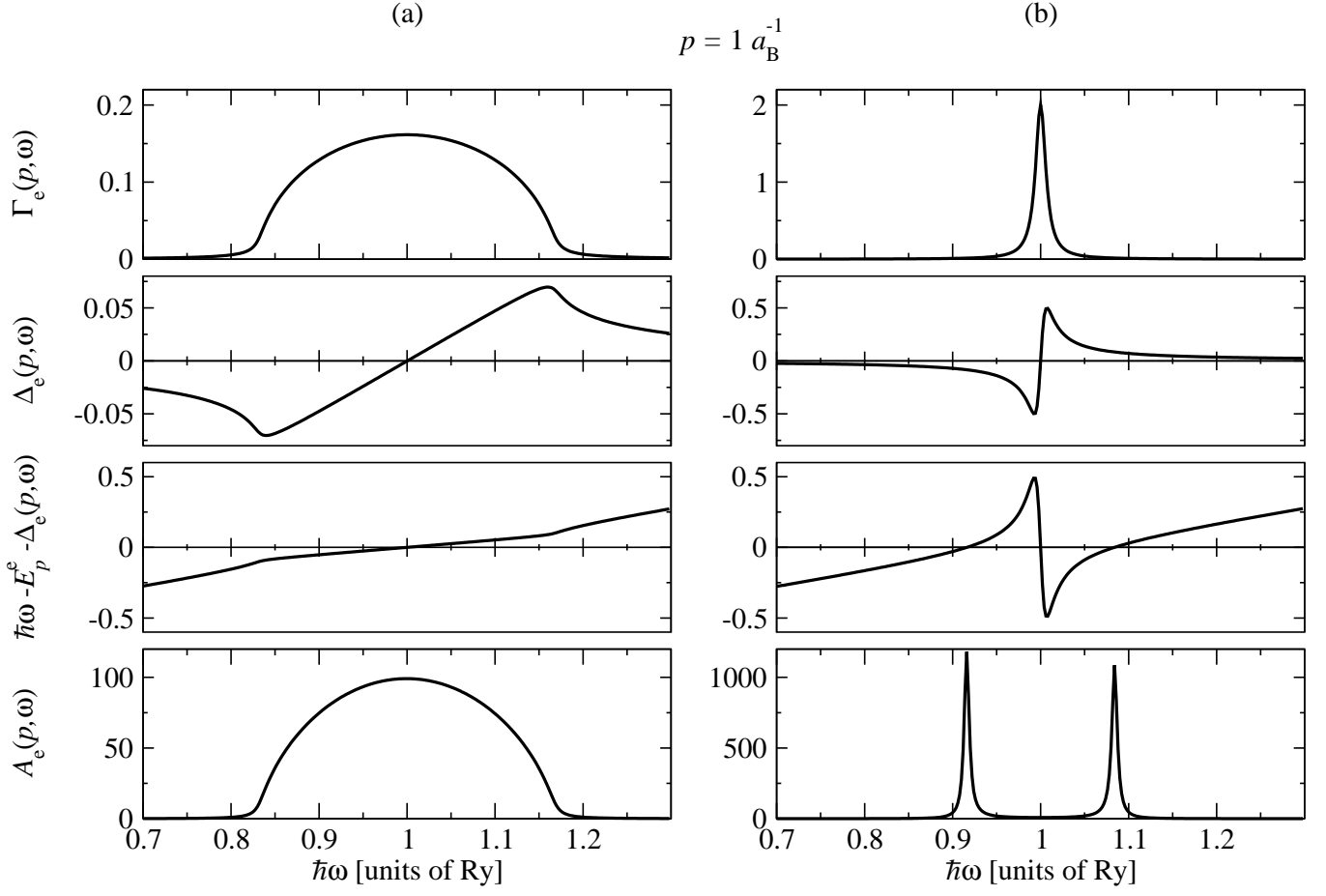}
	\caption{Imaginary and real part of the electrons self-energy 
	$\Sigma_\mathrm{e}(p,\hbar\omega)$, 
	dispersion relation, and spectral function $A_\mathrm{e}(p,\hbar\omega)$ for 
	$p=1~a_\mathrm{B}^{-1}$  (a) from self-consistent calculation cf. Eq.~(\ref{eqn:se_iteration}) and
	(b) using free propagators, cf. Eq.~(\ref{eqn:se_free}). 
	The parameters are:
	Electron density $n_\mathrm{e}=10^{-6}a^{-3}_\mathrm{B}$,  
	temperature $k_\mathrm{B}T=27.2\ \mathrm{eV}$. 
	\label{fig3}}
\end{figure}

From the form of the screened potential Eq.~(\ref{eqn:debye-pot}),
we note that the main contribution to the integral in 
Eq.~(\ref{eqn:se_result}) arises from terms with small momentum $q$. 
Therefore, we will discuss an approximation where the argument $\boldsymbol p+\boldsymbol q$ in the self-energy 
is replaced by 
$\boldsymbol p$, i.e. $\Sigma_\mathrm{e}(\boldsymbol p+\boldsymbol
q,\hbar\omega ) \approx 
\Sigma_\mathrm{e}(\boldsymbol p,\hbar\omega )$ on the r.h.s. 
This ansatz is justified as can be seen in FIG.~\ref{fig2},
where we show the full solution of Eq.~(\ref{eqn:se_result}) and
the approximative solution for the set of parameters $n_\mathrm{e}=10^{-6}\,a_\mathrm{B}^{-3}$ and $ k_\mathrm{B}T=2\ \mathrm{Ry}=27.2\ \mathrm{eV}$. 
After dropping the shift of the
momentum variable in the self-energy, the $q$ integral can be
performed analytically. The result
\begin{eqnarray}\nonumber
        && \Sigma_\mathrm{e}(\boldsymbol p,\hbar\omega+i\eta)= 
        -\frac{n_{\mathrm i}m_\mathrm{e}e^4}{4\pi\epsilon_0^2
        \hbar^2}
        \frac{1}{\kappa}
        \Bigg[\kappa^2+p^2
        -2\frac{m_\mathrm{e}}{\hbar^2}(
        \hbar\omega+i\eta-\Sigma_\mathrm{e}(\boldsymbol
        p,\hbar\omega+i\eta) )
        -2i\kappa\sqrt{2\frac{m_\mathrm{e}}{\hbar^2}(
        \hbar\omega+i\eta -\Sigma_\mathrm{e}(\boldsymbol
        p,\hbar\omega+i\eta) )}
        \Bigg]^{-1}\;,\\
        \label{eqn:se_iteration}
\end{eqnarray}
is solved numerically for the real and imaginary part.
 In FIG.~\ref{fig3} l.h.s. (a), we show the self-energy, the 
dispersion relation, and the resulting spectral function obtained from 
a self-consistent solution of Eq.~(\ref{eqn:se_iteration}).
The spectral function in FIG.~\ref{fig3} shows a broadened
quasi-particle resonance at the energy $\hbar\omega
=\hbar^2(p^2+\kappa^2)/2 m_\mathrm{e}$. As expected, its shape is
primarily determined by the imaginary part of the self-energy. This can be
seen by a comparison of $A_e(p,\omega)$ with $\Gamma_e(p,\omega)$.
        
We mention, that analytic constraints on the self-energy function such as
Kramers-Kronig relations
\begin{equation}
        {\rm Im}\,\Sigma_\mathrm{e}(\boldsymbol p,\hbar\omega )=
        \int\limits_{-\infty}^\infty\frac{\mathrm{d}\omega' }{\pi}
        \frac{ {\rm Re}\,\Sigma_\mathrm{e}(\boldsymbol p,\hbar\omega')}{
        \omega '-\omega }~,\quad
        {\rm Re}\,\Sigma_\mathrm{e}(\boldsymbol p,\hbar\omega )=
        \int\limits_{-\infty}^\infty\frac{\mathrm{d}\omega' }{\pi}
        \frac{ {\rm Im}\,\Sigma_\mathrm{e}(\boldsymbol
          p,\hbar\omega'+i \eta )}{
        \omega '-\omega }\;\;\;,
        \label{eqn:kramers-kronig}
\end{equation}
as well as the first sum-rule for the spectral function
\begin{equation}
        \int\limits_{-\infty}^\infty\frac{\mathrm{d} \hbar\omega }{2\pi}
        A_\mathrm{e}(\boldsymbol p,\hbar\omega)=1\;\;\;, 
        \label{eqn:sumrule}
\end{equation}
are fulfilled within the numerically achievable precision.

For the sake of comparison, we discuss a simplified calculation in which we neglect the self-consistent propagator and
replace it by a free propagator.
Thus, the self-energy on the
r.h.s. of Eq.~(\ref{eqn:sel_diagram}) disappears. Then we find
\begin{eqnarray}
        \Sigma^{0}_e(\boldsymbol p,\hbar\omega+i \eta)
        &=&- \frac{n_{\mathrm i}me^4}{4\pi\epsilon_0^2\hbar^2}
        \frac{1}{\kappa}
        \Bigg[\kappa^2+p^2-2\frac{m_\mathrm{e}}{\hbar^2}(\hbar\omega+i\eta) 
        -2i
        \kappa\sqrt{2\frac{m_\mathrm{e}}{\hbar^2}(\hbar\omega +i\eta)}
        \Bigg]^{-1}~,
        \label{eqn:se_free}
\end{eqnarray}
which can be separated into real and imaginary part in the
limit $\eta \to 0$:
\begin{eqnarray}
        \Delta^0_\mathrm{e}(\boldsymbol p,\hbar\omega )&=&
        -\frac{n_{\mathrm i}me^4}{2(2\pi)^2\epsilon_0^2\hbar^2}
        \frac{\pi}{\kappa}
        \frac{p^2/2+\kappa^2/2-m_\mathrm{e}\omega /\hbar}
        {(p^2/2+\kappa^2/2-m_\mathrm{e}\omega
          /\hbar)^2+2\kappa^2m_\mathrm{e}\omega /\hbar}~, 
        \label{eqn:real_se_born} \\
        \Gamma^0_\mathrm{e}(\boldsymbol p,\hbar\omega )&=& 
        \frac{\pi n_{\mathrm i}me^4}{(2\pi)^2\epsilon_0^2\hbar^2}
        \frac{\sqrt{2m_\mathrm{e}\omega /\hbar}}
        {(p^2/2+\kappa^2/2-m_\mathrm{e}\omega
          /\hbar)^2+2\kappa^2m_\mathrm{e}\omega /\hbar}~. 
        \label{eqn:im_se_born}
\end{eqnarray}
From the imaginary part $\Gamma^0_\mathrm{e}(\boldsymbol p,\hbar\omega )$ we 
see that the  contribution to the spectral function near the 
free-particle energy $\hbar\omega =E^\mathrm{e}_p$ is damped out to a large 
extent.
These functions as well as the corresponding dispersion relation 
$\hbar\omega -E^\mathrm{e}_p-\Delta^0_\mathrm{e}(\boldsymbol p,\hbar\omega )$
and spectral function are 
plotted on the r.h.s. of FIG.~\ref{fig3}. 
The  spectral function exhibits two
separate peaks corresponding to the roots of the dispersion relation
and no peak at the quasi-particle energy $E^\mathrm{e}_p$.
This contribution from 
the central root at the free-particle energy is damped out due to
the large value of $\Gamma^0_\mathrm{e}(\boldsymbol p,\hbar\omega =E^\mathrm{e}_p)$ at the 
same energy. Note also, the order of magnitude change in the damping
$\Gamma_e$ between the first iteration and the self-consistent result FIG.~\ref{fig3} l.h.s. 
Thus, these structures can clearly be identified as artifacts since they 
disappear in the self-consistent calculation.

\subsection{C. Effects of the electron self-energy}
The Born approximation can be improved accounting for self-energy
and vertex corrections, see Eq.~(\ref{eqn:ward}).
However, the contribution of the two diagrams containing the self-energy
is diverging so that we performed partial summations
of higher orders, leading to the Dyson equation and the spectral
function discussed above. In this way, the free electron propagator 
is replaced by the full electron propagator when calculating the force-force correlation function. 
This approximation exactly reflects the
point made by Migdal and Landau/Pomeranchuk to account for a finite
life-time (damping rate) of electron states propagating in a dense
medium.

The force-force Green function is obtained from the evaluation of
the following diagram:\\[1cm]
\begin{eqnarray}
        G^\Sigma_{\dot J\dot J}(\omega_\mu )&=& 
	\parbox{.2\textwidth}{
	\includegraphics[width=.2\textwidth,clip]{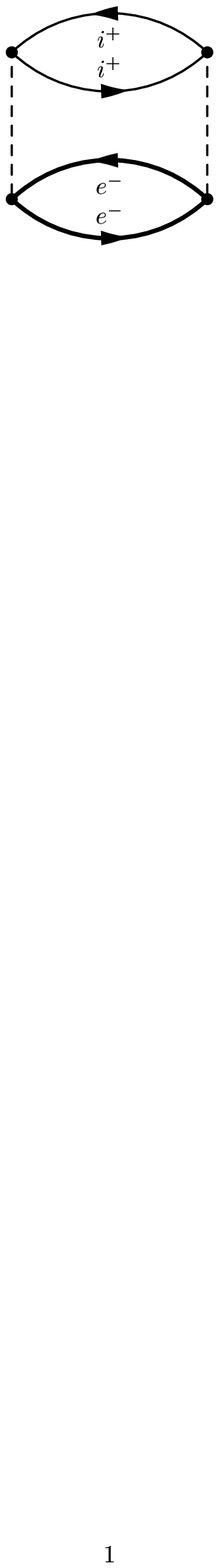}}
       \label{eqn:force_spek}\\\nonumber\\
        &=&
        \frac{n_\mathrm{i}e^6}{m_\mathrm{e}^2\epsilon_0^2}
        \sum\limits_{z_\nu}
        \int \frac{\mathrm{d} ^3p\,}{(2\pi)^3}
        \int\frac{\mathrm{d} ^3q}{(2\pi)^3}
        \frac{1}{(q^2+\kappa^2)^2}q_z^2
        \int\limits_{-\infty}^\infty
        \frac{\mathrm{d} \hbar\omega' }{2\pi }
        \frac{A_e(\boldsymbol p+\boldsymbol q,\hbar\omega ')}{z_\nu+\omega _\mu-\hbar\omega '}
        \int\limits_{-\infty}^\infty
        \frac{\mathrm{d} \hbar\omega'' }{2\pi }
        \frac{A_e(\boldsymbol p,\hbar\omega'')}{z_\nu-\hbar\omega''}.
\end{eqnarray}
After summation over the fermionic Matsubara frequencies $z_\nu$ and
shifting variables,
we obtain the imaginary part of $G^{\Sigma}_{\dot J\dot J}(\omega )$,
cf. also App. A,

\begin{eqnarray}
&&{\rm Im}\,G^\Sigma_{\dot J\dot J}(\omega )= 
        \frac{ n_\mathrm{i}e^6}{6 m_\mathrm{e}^2\epsilon_0^2}
        \int \frac{\mathrm{d} ^3p\,}{(2\pi)^3}
        \int\frac{\mathrm{d} ^3q}{(2\pi)^3} 
        \frac{(\boldsymbol{p} - \boldsymbol{q})^2}{((\boldsymbol{p} - 
        \boldsymbol{q})^2+\kappa^2)^2}
        \int\limits_{-\infty}^\infty
        \frac{\mathrm{d} \hbar\omega' }{2\pi }
        A_e(q,\hbar\omega ')        
        A_e(p,\hbar\omega+\hbar\omega')\left(f(\hbar\omega '+\hbar\omega
          )-f(\hbar\omega ')\right) 
\nonumber\\&&=  
        \frac{\pi n_\mathrm{i}e^6\hbar}
        {m_\mathrm{e}^2\epsilon_0^2(2\pi)^8}
        4\pi\int\limits_{0}^{\infty}\!\!
        \mathrm{d} p\,p^2\,\frac{2\pi}{3}
        \int\limits_{0}^{\infty}\!\! \mathrm{d} q q^2\,
        [-\frac{2 \kappa^2}{((p+q)^2+\kappa^2)((p-q)^2+\kappa^2)}
        +\frac{1}{2pq}
        \ln\left|\frac{(p+q)^2+\kappa^2}{(p-q)^2+\kappa^2}\right|]\times
        \nonumber\\&& \hspace*{4cm} \times
        \int\limits_{-\infty}^\infty\!\!\mathrm{d}\omega'
        A_e(q,\hbar\omega '+\hbar\omega )
        A_e(p,\hbar\omega')
        \left(f(\hbar\omega '+\hbar\omega )-f(\hbar\omega ')\right)~,
        \label{eqn:force_spek_final}
\end{eqnarray}
where the integrals over the angular parts have been performed.
 
The further evaluation requires an expression for the spectral function.
Note, that in the limit of free particles, where the spectral function
is given by a $\delta$-function, the Born approximation, 
Eq. (\ref{eqn:born_110}), is recovered. We will use the result obtained above 
within our approximation for the self-energy.

\begin{figure}
	\includegraphics[width=.75\textwidth,clip,angle=-90]{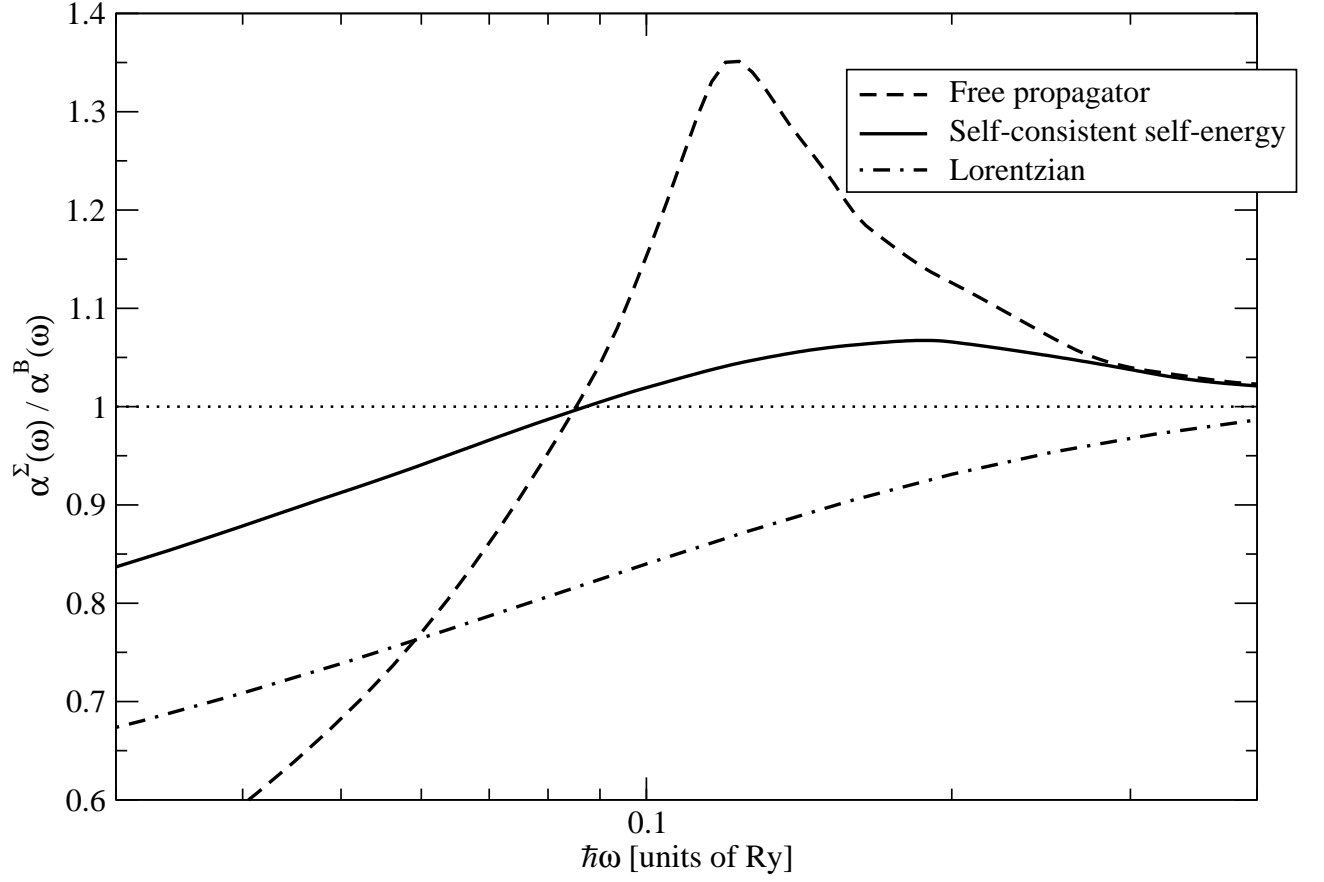}
	\caption{Correction factor $\alpha^\mathrm{\Sigma}(\omega)
	/\alpha^\mathrm{\rm B}(\omega )$ as a function of the photon
	energy $\hbar\omega$
	with free propagators in self-energy diagram, the self-consistent
	spectral function, and a Lorentzian ansatz. Parameter values:
	Electron density $n_\mathrm{e}=10^{-6}a^{-3}_\mathrm{B}$,  
	temperature $k_\mathrm{B}T=27.2\ \mathrm{eV}$ ~.
	\label{fig4}}
\end{figure}

The result is shown in FIG.~\ref{fig4}.
The correction factor $\alpha^{\Sigma}(\omega)/\alpha^{\rm
B}(\omega)$ (cf. Eq.(\ref{eq:alpha_green})) is plotted as a function of the frequency for 
three different approximations with the parameters 
$n_\mathrm{e}=10^{-6}\,a_\mathrm{B}^{-3}$ and $ k_\mathrm{B}T=27.2\ \mathrm{eV}$.
The full line presents the result of the
self-consistent treatment, Eq.~(\ref{eqn:se_iteration}), and is
compared to a calculation with free propagators in the 
self-energy diagrams, see Eq.~(\ref{eqn:ward}), and a calculation 
using a Lorentzian ansatz of the
spectral function with a width taken at the on-shell energy $\Gamma(\boldsymbol{p},\hbar\omega =E^\mathrm{e}_p)$. 
This corresponds to the introduction of a 
finite life-time in the approach of Knoll and Voskresensky 
\cite{knol:annals96}.

For the lowest frequencies considered here, all approximations show
a suppression of the absorption coefficient as compared to the Born
result. At high frequencies, all curves tend towards unity, i.e.,
the Born result is recovered. For intermediate frequencies, 
an enhancement of up to 35 \% 
is found for the calculation using free propagators. Making use of the
self consistency, the enhancement is reduced to 6 \% at most. 
For the Lorentzian ansatz, no enhancement at all appears.
Thus, the width of the imaginary part of the self-energy as a function of frequency $\omega $ (FIG.~\ref{fig3}) 
plays a crucial r\^ole
for the size of the enhancement. We expect that any increase in the 
width of the self-energy will further decrease the enhancement or even
lead to a suppression for all frequencies as in the case of the
Lorentzian ansatz, which corresponds to an infinite 
width of the imaginary part of the self-energy. A further broadening
could result from an extension to higher orders in the
set of diagrams for the self-consistent calculation of the
spectral function, e.g. by inclusion of vertex terms. 

\subsection{D. Vertex corrections}

\begin{figure}
	\includegraphics[width=.8\textwidth,clip,angle=-90]{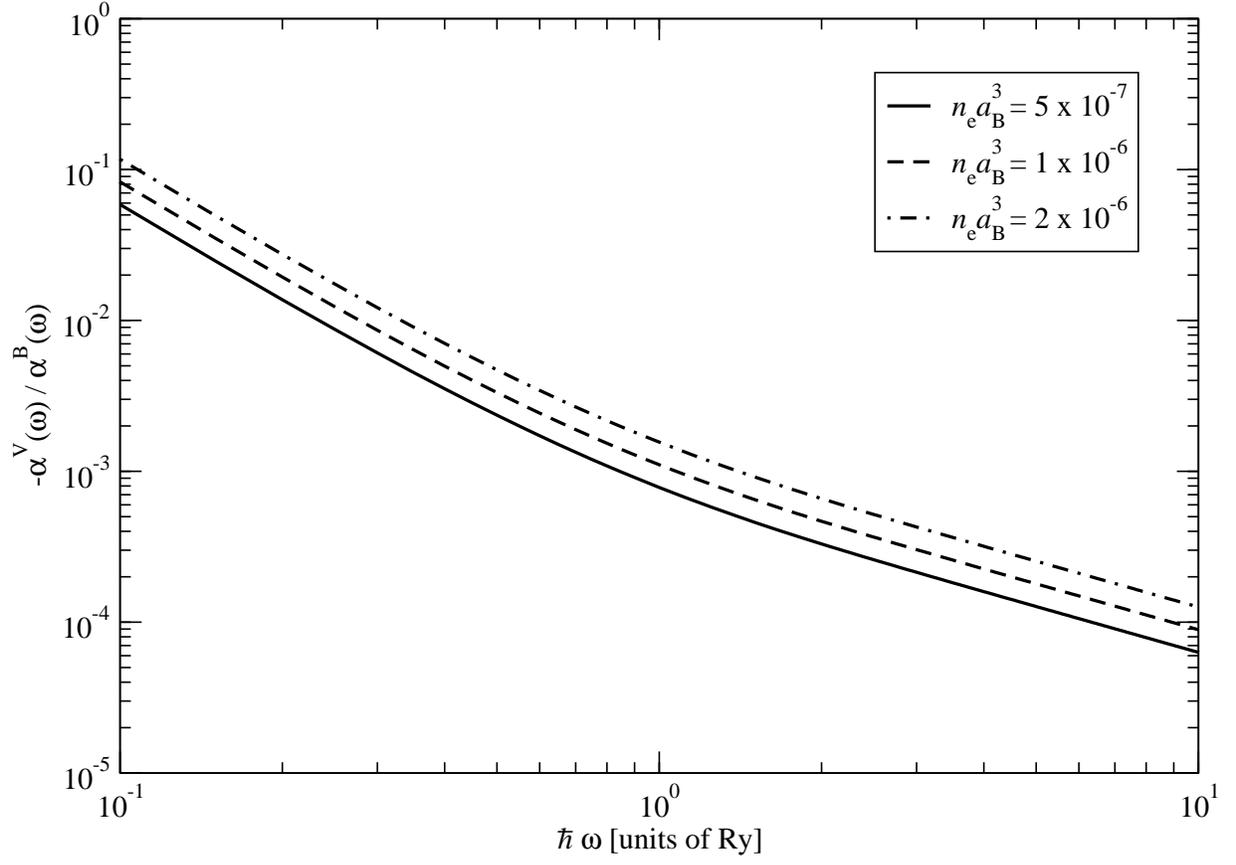}
	\caption{Correction factor $\alpha^\mathrm{V}(\omega )
	/\alpha^\mathrm{\rm B}(\omega )$ as a function of the photon
	energy $\hbar\omega $. For high energies the 
	ratio behaves like $1/\omega $. Various values of the 
	electron density are considered. Temperature: $k_\mathrm{B}T=27.2\ \mathrm{eV}$.
	\label{fig5}}
\end{figure}

\noindent
As known from the Ward-Takahashi identities, self-energy 
and vertex corrections are intrinsically related. In particular, if
medium corrections arise in a certain order of a small parameter like
the density from the self-energy, also contributions from the vertex
corrections are expected in the same order. This well-known fact is used to
construct so-called conserved approximations \cite{kada:pr61,baym:pr62} and 
has also been discussed in the
context of other medium effects, such as the  modification of
two-particle states or the inclusion of bound states into the
polarization function and the calculation of optical line spectra
profiles \cite{guen:habil}. Thus, it is necessary to study the 
vertex corrections corresponding to the self-energy considered before.
However, the solution of the vertex-equation is a technically very  challenging
task and has been solved so far only in certain approximations \cite{takada,maha:prb94}.
In lowest order, the vertex correction is obtained through insertion of one 
ion-loop inside the electron-loop:\\[.5cm]
\begin{equation}
        \label{eq:gjj_v}
        G_{\dot J\dot J}^{\rm V}(\omega _\mu)=\ 
	\parbox{.2\textwidth}{
	\includegraphics[width=.2\textwidth,clip]{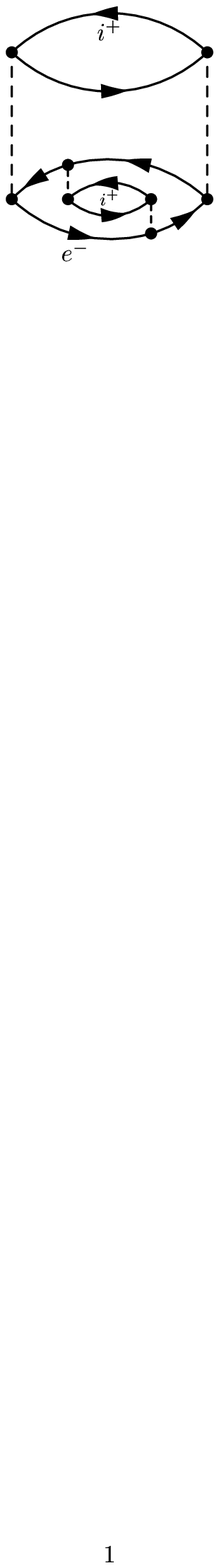}} ~.
        \end{equation}
\\[1cm]
We find for the imaginary part of the force-force Green function
   \begin{eqnarray} {\rm Im}\,G^{\mathrm V}_{\dot J\dot J}(\omega+i\eta)&=&
        \frac{ n_\mathrm{i}
        n_\mathrm{e} \Lambda_\mathrm{e}^3 e^6}{24 \pi^3\epsilon_0^2 \hbar^3 
        m_\mathrm{e} \omega}
        \bigg[\, \int\limits_0^\infty  \mathrm{d}p\,p\, 
        e^{-\beta \hbar^2 p^2/2m_\mathrm{e}} \Gamma^{\mathrm V}(p) 
	\ln\left|\frac{\sqrt{E^\mathrm{e}_p} + \sqrt{E^\mathrm{e}_p+\hbar\omega }
        }{\sqrt{E^\mathrm{e}_p} - \sqrt{E^\mathrm{e}_p+\hbar\omega }} 
        \right|+ 
        \nonumber\\
	\label{eqn:vert_green}
        &&
        \qquad\qquad\ +
        \int\limits_{\sqrt{2 m_\mathrm{e} \omega / \hbar}}^\infty  
        \!\!\!\!\!\!\!\mathrm{d} p\,p\, {\rm e}^{-\beta \hbar^2
	p^2/2m_\mathrm{e}}\Gamma^{\mathrm V}(p)\ln\left|\frac{\sqrt{E^\mathrm{e}_p} + \sqrt{E^\mathrm{e}_p-\hbar\omega }}
	{\sqrt{E^\mathrm{e}_p} - \sqrt{E^\mathrm{e}_p-\hbar\omega }} 
        \right| \bigg] ~,
\end{eqnarray}
with the vertex part
\begin{eqnarray}
        \Gamma^{\mathrm V}(p) &=&  {\pi^2 n_{\mathrm i} e^4 m \over
          (2 \pi)^3 \hbar^2 
  \epsilon_0^2 \kappa p^2}~, 
\end{eqnarray}
taken in lowest order in $\kappa$. 
For the details, see App. B.
This expression can be evaluated numerically. In the limit of 
high frequencies, where the second 
integral becomes negligible compared to the first, the absorption 
coefficient is proportional to $K_1(\omega )/\omega^4 $. 
Since the $K_1$ function has the same asymptotic behavior as the
$K_0$ function, which was the characteristic of the Born result for
the absorption coefficient Eq.~(\ref{eq:alpha_born}), the ratio 
$\alpha^{\mathrm V}/\alpha^{\rm B}$ (cf. Eq.~(\ref{eq:alpha_green}))
behaves like $1/\omega $ in the high frequency limit.
The relative change of the absorption coefficient due to the
vertex correction $\alpha^{\rm V}/\alpha^{\rm B}$ is shown in FIG.~\ref{fig5}.
        
For all frequencies considered, a suppression with respect to the
Born approximation is found.
For the considered energy range, the corrections are small and
decrease with increasing energy. For energies larger than 1 Ry, the
expected high frequency behavior $\propto\omega^{-1}$ arises. 
The corrections are small for low densities compared to higher
densities. 

\begin{figure}
	\includegraphics[width=.8\textwidth,clip,angle=-90]{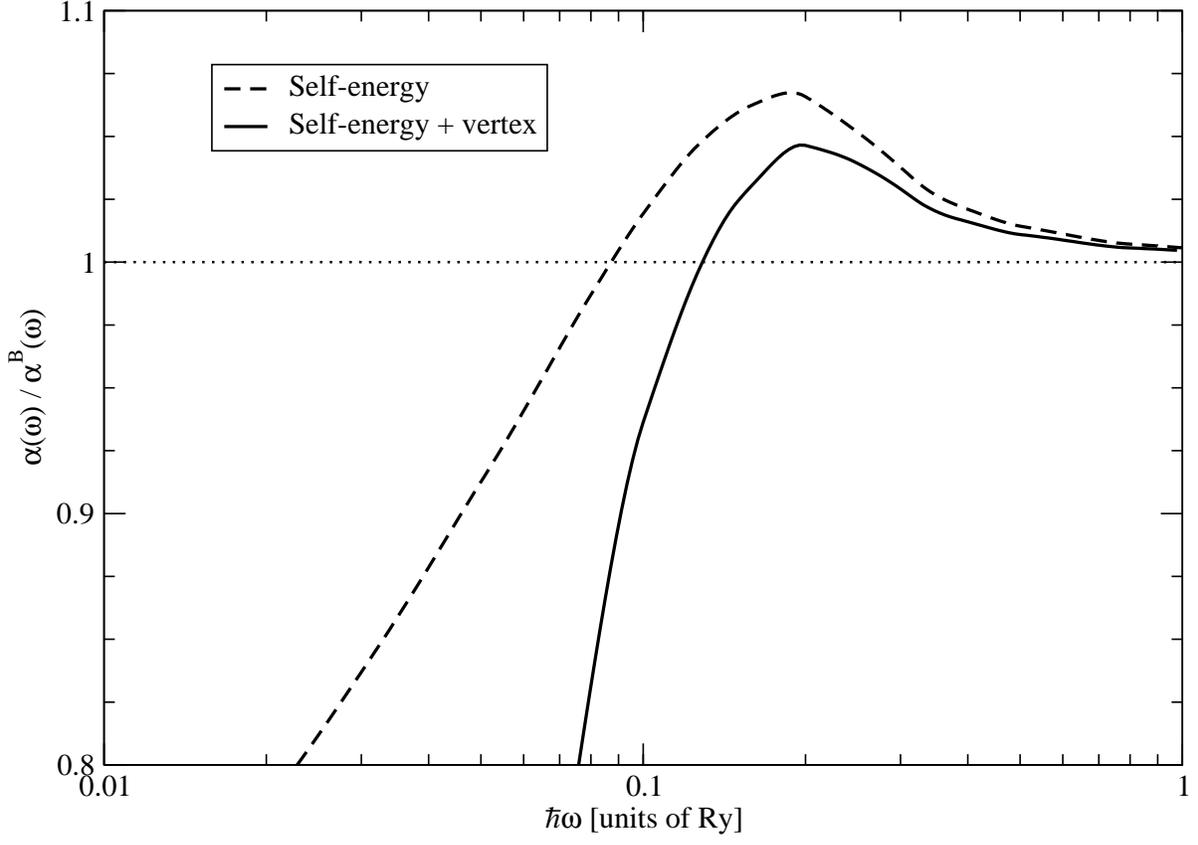}
	\caption{Total absorption coefficient taken relative to the 
	Born result $\alpha^\mathrm{V}(\omega )
	/\alpha^\mathrm{\rm B}(\omega )$ as a function of the photon
	energy $\hbar\omega$. Parameter values:
	Electron density $n_\mathrm{e}=10^{-6}a^{-3}_\mathrm{B}$,  
	temperature $k_\mathrm{B}T=27.2\ \mathrm{eV}$.
	\label{fig6}}
\end{figure}

We consider the absorption coefficient
$\alpha(\omega)$ including all of the improvements.
Since the Born approximation is already included in the self-energy contribution $\alpha^\Sigma(\omega )$ we have
\begin{eqnarray}
  \alpha(\omega) & = & 
  \alpha^{\rm V}(\omega)\,+
\alpha^{\Sigma}(\omega) 
\;\;\;.
\end{eqnarray}
The relative change $\alpha(\omega)/\alpha^{\rm B}(\omega)$ 
is presented in FIG.~\ref{fig6}. For the sake of comparison,
the self-energy correction is shown as well. For small frequencies,
the self-energy contribution and the vertex contribution add to
a net suppression. For higher energies, the self-energy term shows
an enhancement, which is partially compensated by the vertex.
However, the net result is still an enhancement. 
In the high frequency limit, the Born result is
reproduced.

\section{4. CONCLUSIONS}

In this paper, we have studied the influence of 
the surrounding medium on the
bremsstrahlung spectrum in non-ideal plasmas. The interaction
with the medium leads to a finite life-time of the electron
states. Instead of free quasi-particles, the spectral function
has to be used to describe the electron properties in the 
medium. Thus, the use of the single-particle spectral function is a 
quantum-statistically sound implementation of the original idea of 
successive scatterings by Landau/Pomeranchuk and Migdal.

Our approach, namely the microscopic treatment of the dynamical self-energy,  
extends a recent work of Knoll and Voskresensky
\cite{knol:annals96}, where a Lorentzian ansatz for the spectral
function with a frequency-independent quasi-particle lifetime was considered.
The Lorentzian ansatz for the spectral function was discussed above in
subsection 3.C., taking the imaginary part of the self-energy at the 
quasi-particle energy. 
Then, a suppression of the bremsstrahlung spectrum was observed.
In general, the microscopical treatment leads a frequency-dependent 
imaginary part of the self-energy, and, according to the
Kramers-Kronig relation, to a non-vanishing real part. In particular, the
inclusion of the real part of the self-energy in the spectral function
influences the medium modification of the bremsstrahlung spectrum.

In the present paper, the one-loop approximation was taken for the
self-energy. It has been shown that a self-consistent treatment has to be
used in order to avoid unphysical artifacts which arise, if instead of the
full propagator the free propagator is taken to evaluate the
self-energy. 
This is already known from the treatment of the 
spectral functions in
plasma physics in the so-called GW approximation
\cite{maha} where the interaction with the medium is implemented by a
screened potential. It should be mentioned that in this case a
self-consistent treatment of the spectral function on the
level of the GW approximation has been performed \cite{WR}. Any
iterative solution starting from the free propagator leads to
non-physical structures in the spectral function, see also \cite{fehr}.

Within our approach, we found a switch from a suppression at low frequencies to
an enhancement at high frequencies for the bremsstrahlung spectrum.
The approximation for the self-energy can be improved by considering
further diagrams. In particular, the vertex correction would be of
interest which modifies the coupling to the interaction
potential. The inclusion of vertex corrections is also necessary to obtain
conserved approximations and has been shown in Subsection 3.D., where a
further suppression of the bremsstrahlung spectrum was observed. In conclusion,
the modification of the bremsstrahlung spectrum by the surrounding medium is
sensitively dependent on the approximation used.

We cannot elaborate further on the switch
from suppression to enhancement seen in the self-energy correction.
In order to verify the existence of such a switch, 
higher order calculations are necessary. A consistent procedure would consist in
a) using the full propagator in the calculation of the vertex correction,
b) solving the full vertex equation with full propagators and finally c) solve
the Dyson equation for the single particle propagator with the solution of the
vertex equation.

The importance of vertex corrections in the self
consistency relations has also been shown in the description of the
spectral function of the homogeneous electron gas. There, notable differences between a so-called
GW$\Gamma$ approximation \cite{takada}
including vertex terms and a GW approximation \cite{holm}
arise. It should be mentioned that
self-consistent Schwinger-Dyson equations for the
self-energy have been considered in field theory \cite{Craig} to find
solutions for the QCD running coupling problem. A corresponding treatment
would lead to a better description of the modification of the
bremsstrahlung spectrum in a dense medium, but would exceed the frame of the present
work.
Also, at low frequencies, further effects such as
the dielectric suppression are of importance.
It has not been considered in this approach, but can easily be obtained
from the force-force correlation function as well.

\acknowledgments{
\section{ACKNOWLEDGEMENTS}

We would like to thank J. Knoll, D. Voskresensky and V. Morozov for stimulating discussions.
C.F. would also like to thank the Gesellschaft f\"ur 
Schwerionenforschung (GSI) for its hospitality 
and the Studienstiftung des Deutschen Volkes for a scholarship.
}

\begin{appendix}

\section{APPENDIX A: Details on the Born approximation}

\renewcommand{\theequation}{A\arabic{equation}}
\setcounter{equation}{0}

The four-particle Green function $G_{\dot J \dot J}$ in Born 
approximation is given by
\begin{eqnarray}
        G^\mathrm{Born}_{\dot J \dot J}(\omega_{\mu})  &=& 
        \label{eq:green_born}
        \frac{e^2}{m_\mathrm{e}^2 \Omega_0^2} 
	\sum_{\boldsymbol{pkq}\atop \omega _\lambda,z_\nu,z_{\bar\nu}}
        \left| v_q^{\rm ei}\right|^2
        q_z^2 \frac{1}{z_{\nu}-E_k^{\rm i}}\,\frac{1}
        {z_{\nu}-\omega_{\lambda}
        -E_{ k- q}^{\rm i}}\,
        \frac{1}{z_{\bar \nu}-E^{\rm e}_p}\,
        \frac{1}{z_{\bar \nu}+\omega_{\mu}+
        \omega_{\lambda}-E_{ p+ q}^{\rm e}} \;\;\;.
\end{eqnarray}
Here, $z_{\nu},z_{\bar \nu}$ denote Fermionic Matsubara frequencies,
while $\omega_{\mu},\omega_{\lambda}$ are Bosonic Matsubara 
frequencies. We use the convention 
$z_\nu=(2\nu+1)\pi i/\beta+\mu_c,\, \nu=0,\pm 1,\pm 2,\ldots$ 
and
$\omega_\lambda=2\lambda\pi i/\beta,\, \lambda=0,\pm 1,\pm 2,\ldots$. 
$\mu_c$ is the chemical potential of a particle of species $c$.
We refer to Ref.~\cite{krae} for details concerning thermodynamic Green
functions. The summation over $z_{\nu},z_{\bar
\nu},\omega_{\lambda}$ can be carried out analytically.

In the adiabatic approximation the summation over $z_{\bar{\nu}}$ and $k$ 
gives the ionic density $n_\mathrm{i}$. 
After summation of $z_\nu$ we have
\begin{equation}
        G^\mathrm{Born}_{\dot J \dot J}(\omega_\mu) =  
        {n_{\mathrm i}e^2 \over m_\mathrm{e}^2
	\Omega_0} \sum_{\boldsymbol{pq}}|v^\mathrm{ei}_q|^2 q_z^2  
	\frac{f(E^\mathrm{e}_p)-f(E^\mathrm{e}_{p+q})}{E^\mathrm{e}_p +\omega_\mu - 
        E^\mathrm{e}_{p+q}}~,
        \label{born_full}
\end{equation} 
and by virtue of the Dirac identity we obtain after analytic continuation
\begin{equation}
        \label{eqn:born_delta}
        {\rm Im}\,G^\mathrm{Born}_{\dot J \dot J}(\omega+ i  \eta) = 
        -{\pi n_{\rm i}e^2 \over m_\mathrm{e}^2
	\Omega_0} \sum_{\boldsymbol{pq}}|v^\mathrm{ei}_q|^2 q_z^2 
        \left[f(E^\mathrm{e}_p)-f(E^\mathrm{e}_{p+q})\right]
        \delta( \hbar \omega +E^\mathrm{e}_p - E^\mathrm{e}_{p+q}) \ .
\end{equation} 
Replacing the sums over momenta by integrals and 
assuming a Maxwellian plasma,
we find
\begin{eqnarray}
        {\rm Im}\,G^\mathrm{Born}_{\dot J\dot J}&=& 
        \frac{\pi n_\mathrm{i} n_\mathrm{e} \Lambda_\mathrm{e}^3e^2 }{
        m_\mathrm{e}(2\pi)^6 } 
        \frac{1 - \mathrm {\rm e} ^{-\beta \hbar \omega} }{(\hbar \omega)^2} 
        \int\limits_{0}^\infty \mathrm{d}p\, p\, 4 \pi\,
        \mathrm{e}^{-\beta \hbar^2 p^2/2m_\mathrm{e}}\!\!
        \int\limits_{0}^\infty \mathrm{d}q\, 2 \pi
        {q^3 \over 3}
        \left( {e^2  \over \epsilon_0 \Omega_0  
        (q^2 + \kappa^2)} \right)^2
        \int\limits_{-1}^1 \mathrm{d} z\, \delta(z - {m_\mathrm{e} 
        \omega \over \hbar pq} + 
        {q \over 2p})~.\nonumber\\\label{eqn:born_101}
\end{eqnarray}
The limits of the $q$-integration are obtained from the
root of the argument of the delta function, 
$p \ge \left|{m_\mathrm{e} \omega \over  \hbar q} - {q \over 2}
\right|$ or equivalently $-p + 
\sqrt{p^2 + 2 m_\mathrm{e} \omega/ \hbar} \le q \le
p +  \sqrt{p^2 + 2 m_\mathrm{e} \omega/ \hbar}$. 
Performing the integrations first over $q$ and then over $p$, we 
obtain the result Eq.(\ref{eqn:born_110}).

\section{APPENDIX B: Details on the vertex correction}

\renewcommand{\theequation}{B\arabic{equation}}
\setcounter{equation}{0}

The vertex contribution to  the force-force Green function is given
by
\begin{eqnarray}
        &&G^{\mathrm V}_{\dot J \dot J}(\omega_\mu) = 
        {e^2 \over m^2 \Omega_0^2} 
	\sum_{\boldsymbol{pkqk'q'}\atop z_\nu z_\mu z_\mu'
        \omega_\lambda  \omega_\lambda'}|v_q^{ \rm ei}|^2 q_z^2 
        |v^{\rm ei}_{q'}|^2{1 \over z_\mu - E^\mathrm{i}_k} {1
        \over z_\mu - \omega_\lambda - E^\mathrm{i}_{k-q}} 
        {1 \over  z_\nu - E^\mathrm{e}_p} 
        {1 \over z_\nu + \omega_\mu 
        + \omega_\lambda - E^\mathrm{e}_{p+q}}\times
        \nonumber\\ &&\qquad\qquad \qquad\times
        {1 \over z_\mu' - E^\mathrm{i}_{k'}}{1 \over z_\mu' - 
        \omega_\lambda' - E^\mathrm{i}_{k'-q'}}
        {1 \over z_\nu +
        \omega_\mu + \omega_\lambda  + \omega_\lambda'
        - E^\mathrm{e}_{p+q+q'}}
        {1 \over z_\nu + \omega_\lambda' - 
        E^\mathrm{e}_{p+q'}}~.
\end{eqnarray}

Again, summation over ionic variables $z_\mu, z'_{\mu}, k, k'$ and 
bosonic frequencies $\omega_{\lambda}$ and $\omega'_{\lambda}$ can be
performed, which gives in adiabatic approximation the ion density $n_{\mathrm i}$. 
Thus,
\begin{eqnarray}
       G^{\mathrm V}_{\dot J \dot J}(\omega_\mu) = 
       \frac{e^2}{m^2\Omega^2_0}
       n_{\mathrm i}^2 \sum_{\boldsymbol{pqq'}, z_\nu} 
        |v_q^{ \rm ei}|^2 q_z^2 |v^\mathrm{ei}_{q'}|^2
        {1 \over z_\nu - E^\mathrm{e}_p} {1 \over
        z_\nu + \omega_\mu - E^\mathrm{e}_{p+q}}
        {1 \over z_\nu + \omega_\mu  - E^\mathrm{e}_{p+q+q'}}
        {1 \over z_\nu - E^\mathrm{e}_{p+q'}}~.
\end{eqnarray} 

Expansion into partial fractions with 
respect to $z_\nu$ and summation leads to
\begin{eqnarray}
        G^{\mathrm V}_{\dot J \dot J}(\omega_\mu) &=&  
	n_{\mathrm i}^2 \sum_{\boldsymbol{pqq'}} |v_q^{ \rm ei}|^2 q_z^2 
        |v^\mathrm{ei}_{q'}|^2
        \Big\{ f(E^\mathrm{e}_{p+q}) {1 \over  E^\mathrm{e}_{p+q} - 
        \omega_\mu - E^\mathrm{e}_p} 
        {1 \over  E^\mathrm{e}_{p+q} -
        \omega_\mu - E^\mathrm{e}_{p+q'}} 
        {1 \over  E^\mathrm{e}_{p+q} - E^\mathrm{e}_{p+q+q'}} 
        \nonumber\\ && 
        +f(E^\mathrm{e}_{p}) {1 \over  E^\mathrm{e}_{p} + \omega_\mu - 
        E^\mathrm{e}_{p+q}} 
        {1 \over  E^\mathrm{e}_{p} + \omega_\mu - E^\mathrm{e}_{p+q+q'}}
        {1 \over  E^\mathrm{e}_{p}- E^\mathrm{e}_{p+q'}} +
        \nonumber\\ && 
        +f(E^\mathrm{e}_{p+q'}) {1 \over  E^\mathrm{e}_{p+q'} +
        \omega_\mu - E^\mathrm{e}_{p+q}} 
        {1 \over  E^\mathrm{e}_{p+q'} + \omega_\mu -  
        E^\mathrm{e}_{p+q+q'}}
        {1 \over  E^\mathrm{e}_{p+q'} - E^\mathrm{e}_p} 
        \nonumber\\ && 
        +f(E^\mathrm{e}_{p+q+q'}) 
        {1 \over  E^\mathrm{e}_{p+q+q'} - \omega_\mu - E^\mathrm{e}_p} 
        {1 \over  E^\mathrm{e}_{p+q+q'} - \omega_\mu - 
        E^\mathrm{e}_{p+q'}}  
        {1 \over  E^\mathrm{e}_{p+q+q'} -  E^\mathrm{e}_{p+q}} \Big\}~.
        \label{eqn:vertex_30}
\end{eqnarray} 
Rigorously, one would have to perform another expansion into 
partial fractions in order to obtain the imaginary part of this 
Green function. This procedure would lead to delta functions of the
form $\delta(E^\mathrm{e}_p-E^\mathrm{e}_{p+q})$ without a bosonic 
frequency of the 
external field in the argument. These are residues of a perturbation
expansion of the wave functions of the electron in the ion's 
potential and do not yield any information about the dynamics of
the system. Thus, only those denominators in Eq.(\ref{eqn:vertex_30}) 
are of interest, that do contain the frequency $\omega_\mu$. It 
suffices to perform an expansion into partial fractions with
respect to $\omega_\mu$. We obtain
\begin{eqnarray}
        \mathrm{Im}\,G^{\mathrm V}_{\dot J \dot J}(\omega) &=&   
	n_{\mathrm i}^2 \pi \sum_{\boldsymbol{pqq'}} |v_q^{
        \rm ei}|^2 q_z^2 |v^\mathrm{ei}_{q'}|^2
        \Big[- f(E^\mathrm{e}_{p+q})\  
        \delta(\hbar\omega + E^\mathrm{e}_p - E^\mathrm{e}_{p+q})  
        {1 \over  E^\mathrm{e}_{p} - E^\mathrm{e}_{p+q'}} 
        {1\over E^\mathrm{e}_{p+q}-E^\mathrm{e}_{p+q+q'}}-\nonumber\\&& 
        - f(E^\mathrm{e}_{p+q})\  \delta(\hbar\omega + E^\mathrm{e}_{p+q'} - 
        E^\mathrm{e}_{p+q}) {1 \over  E^\mathrm{e}_{p+q'}-
        E^\mathrm{e}_{p}} {1 \over  E^\mathrm{e}_{p+q} - 
        E^\mathrm{e}_{p+q+q'}} +
        \nonumber\\ && +
        f(E^\mathrm{e}_{p})\  \delta(\hbar\omega + E^\mathrm{e}_p - 
        E^\mathrm{e}_{p+q}) {1 \over  E^\mathrm{e}_{p} -
        E^\mathrm{e}_{p+q'}} 
        {1 \over  E^\mathrm{e}_{p+q} - E^\mathrm{e}_{p+q+q'}}+ 
        \nonumber\\ && 
        +f(E^\mathrm{e}_{p})\  \delta(\hbar\omega +
        E^\mathrm{e}_p - E^\mathrm{e}_{p+q+q'})  
        {1 \over  E^\mathrm{e}_{p} - E^\mathrm{e}_{p+q'}} 
        {1 \over  E^\mathrm{e}_{p+q+q'} -
        E^\mathrm{e}_{p+q}}+ 
        \nonumber\\ && +
        f(E^\mathrm{e}_{p+q'})\  \delta(\hbar\omega + E^\mathrm{e}_{p+q'} - 
        E^\mathrm{e}_{p+q}) {1 \over  E^\mathrm{e}_{p+q'} -
        E^\mathrm{e}_{p}} 
        {1 \over  E^\mathrm{e}_{p+q} - E^\mathrm{e}_{p+q+q'}} +
        \nonumber\\ && 
        + f(E^\mathrm{e}_{p+q'})\  
        \delta(\hbar\omega+E^\mathrm{e}_{p+q'} - E^\mathrm{e}_{p+q+q'}) 
        {1 \over  E^\mathrm{e}_{p+q'} - E^\mathrm{e}_{p}} 
        {1 \over  E^\mathrm{e}_{p+q+q'} - E^\mathrm{e}_{p+q}}-
        \nonumber\\ && -
        f(E^\mathrm{e}_{p+q+q'})\  
        \delta(\hbar\omega + E^\mathrm{e}_p - E^\mathrm{e}_{p+q+q'})
        {1 \over  E^\mathrm{e}_{p+q+q'} - E^\mathrm{e}_{p+q}} 
        {1 \over  E^\mathrm{e}_{p} - E^\mathrm{e}_{p+q'}}- 
        \nonumber\\ && -
        f(E^\mathrm{e}_{p+q+q'})\  
        \delta(\hbar\omega+E^\mathrm{e}_{p+q'} - E^\mathrm{e}_{p+q+q'}) 
        {1 \over E^\mathrm{e}_{p+q+q'} - E^\mathrm{e}_{p+q}} 
        {1 \over  E^\mathrm{e}_{p+q'} - E^\mathrm{e}_{p}} \Big] ~.
\end{eqnarray}
This can be cast into the form
\begin{eqnarray}  
	&&{\rm Im}\,G^{\mathrm V}_{\dot J \dot J}(\omega+ i \eta)        = n_{\mathrm i}^2 \pi \sum_{\boldsymbol{pqq'}} |v_q^{ \rm ei}|^2 q_z^2
        |v^\mathrm{ei}_{q'}|^2 f(E^\mathrm{e}_{p})
        {1 \over  E^\mathrm{e}_{p} -  E^\mathrm{e}_{p+q'}} 
        {1 \over  E^\mathrm{e}_{p+q} - E^\mathrm{e}_{p+q+q'}} \times
        \nonumber\\ && 
        \times\left[\delta(\hbar\omega+E^\mathrm{e}_p-E^\mathrm{e}_{p+q})-  
        \delta(\hbar\omega + E^\mathrm{e}_p - E^\mathrm{e}_{p+q+q'}) - 
        \delta(\hbar\omega -E^\mathrm{e}_p+E^\mathrm{e}_{p+q})+
        \delta(\hbar\omega-E^\mathrm{e}_p+ E^\mathrm{e}_{p+q+q'}) \right]~,
        \label{eqn:vertex_040} 
\end{eqnarray} 
where only the distribution function $f(E^\mathrm{e}_p)$ appears. 
\\ For small internal momenta $q'$ the denominator 
$ E^\mathrm{e}_{p+q} - E^\mathrm{e}_{p+q+q'}$
can be expanded and 
vanishes after rewriting the delta function as 
$\delta(\hbar\omega + E^\mathrm{e}_p - E^\mathrm{e}_{p+q+q'}) = 
\delta(\hbar\omega + E^\mathrm{e}_p -  E^\mathrm{e}_{p+q} +  
E^\mathrm{e}_{p+q} - E^\mathrm{e}_{p+q+q'})$. 
For $\omega \to 0$, this argument does not hold, but it is used here
in order to find the influence of the vertex correction at 
finite frequencies $\omega$.\\
Eq.~(\ref{eqn:vertex_040}) becomes
\begin{eqnarray}
        {\rm Im}\,G^{\mathrm V}_{\dot J\dot J}(\omega+  i \eta) 
        &=&  {\mathrm{d}  \over \mathrm{d}  \omega} 2 \,n_{\rm i}^2 \pi
	\sum_{\boldsymbol{pqq'}} |v_q^{ \rm ei}|^2 q_z^2  
        |v^\mathrm{ei}_{q'}|^2 f(E^\mathrm{e}_{p}){1 \over  E^\mathrm{e}_{p} -  
        E^\mathrm{e}_{p+q'}} \nonumber\times\\ && 
        \times \left[  
        \delta(\hbar\omega + E^\mathrm{e}_p - E^\mathrm{e}_{p+q})  - 
        \delta(\hbar\omega - E^\mathrm{e}_p + E^\mathrm{e}_{p+q}) \right]~.
\end{eqnarray}
Replacing the summation over the momenta by an integration, 
we have 
\begin{eqnarray}
        &&{\rm Im}\,G^{\mathrm V}_{\dot J\dot J}(\omega+ i  \eta) = 
        \frac{\mathrm{d} }{\mathrm{d} \omega } 
        {2 n_{\mathrm i}^2 \pi \over \hbar^2}\int
        { \mathrm{d} ^3 p \over (2 \pi)^3} \int {\mathrm{d}^3q\over
        (2 \pi)^3} \int {\mathrm{d}^3 q' \over (2 \pi)^3} 
        |v_q^{\rm ei}|^2 q_z^2
        |v^\mathrm{ei}_{q'}|^2 f(E^\mathrm{e}_{p})
        {m_\mathrm{e} \over \boldsymbol p\cdot\boldsymbol q'+q'^2/2}
        \times\nonumber\\ && 
        \times\{\delta(\hbar\omega-\boldsymbol p\cdot
        \boldsymbol q/m_\mathrm{e} -  q^2/2m_\mathrm{e}) -
        \delta(\hbar\omega  +  \boldsymbol p \cdot 
        \boldsymbol q/m_\mathrm{e} +  q^2/2m_\mathrm{e}) \} ~.
        \label{eqn:vertex_50}
\end{eqnarray}
Comparing Eq.(\ref{eqn:vertex_50}) to
the Born result Eq.(\ref{eqn:born_101}),
an additional function arises 
\begin{eqnarray}
        \Gamma^{\mathrm V}(p) &=&  
        {2 n_{\mathrm i} \Omega_0^2 \over (2 \pi)^3 \hbar^2} 
        \int \mathrm{d} ^3q' |v^\mathrm{ei}_{q'}|^2 
        {m_\mathrm{e} \over \boldsymbol p \cdot \boldsymbol q' + 
        q'^2/2} = {4 \pi e^4 n_{\mathrm i} m_\mathrm{e}
        \over (2 \pi)^3 \hbar^2 \epsilon_0^2 p} \int_0^{\infty}\! \mathrm{d} 
        q' {q' \over (q'^2+\kappa^2)^2} 
        \int\limits_{-1}^1\! \mathrm{d}z{1\over z+q'/2p}\nonumber\\&=& 
        \label{vertex_70}
        {4 \pi n_{\mathrm i} e^4 m_\mathrm{e} \over (2 \pi)^3 
        \hbar^2 \epsilon_0^2p} \int_0^{\infty}\! \mathrm{d} q' 
        {q' \over (q'^2+\kappa^2)^2} \ln\left|{1 + q'/2p \over 1 - 
        q'/2p}\right| = 
        {4 \pi^2 n_{\mathrm i} e^4 m_\mathrm{e} \over (2 \pi)^3 
        \hbar^2 \epsilon_0^2 } {1 \over 4 p^2 \kappa + \kappa^3}
\;\;\;,
\end{eqnarray}
which is obtained from the integration over $q'$.
The $q$ integration can be performed as above, cf. Eq.~(\ref{eqn:born_110}).
Insertion into the force-force correlation function 
Eq.(\ref{eq:corr_green}) gives 
\begin{eqnarray}
        \label{vertex_80}
        \big\langle \,\,\dot{\!\!J}_{0}^{z} , \,\,\dot{\!\!J}_{0}^{z} 
        \big\rangle_{\omega +i \eta}^{\rm V} &=&  
        {i  e^2 \hbar \pi \over \beta 
        \Omega_0^3 m_\mathrm{e}^2} \int {\mathrm{d} 
        \hbar \bar\omega \over \pi} 
        {1 \over \omega  - \bar\omega+i \eta} 
        {1 \over \hbar\bar\omega} {\mathrm{d}  \over \mathrm{d}  
	\hbar\bar\omega} \sum_{\boldsymbol{pq}} |v_q^{ \rm ei}|^2 {q^2 \over 3} 
        f(E^\mathrm{e}_p) n_{\rm i} \Gamma^{\rm V}(p)\times\nonumber\\&&\qquad\times \left[\delta (\hbar\bar \omega - 
        {\hbar^2 p q z \over m_\mathrm{e}} - 
        {\hbar^2 q^2 \over 2 m_\mathrm{e}}) + \delta 
        (-\hbar\bar\omega - {\hbar^2 p q z \over m_\mathrm{e}} -
        {\hbar^2 q^2 \over 2 m_\mathrm{e}})\right]~.
\end{eqnarray}
Through partial integration we obtain the simple p\^ole-structure
\begin{equation}
        \label{eqn:vertex_90}
        \big\langle \,\,\dot{\!\!J}_{0}^{z} , \,\,\dot{\!\!J}_{0}^{z} 
        \big\rangle_{\omega+i \eta}^{\rm V} = 
        {i  e^2 \hbar \over \beta \Omega_0^3 m_\mathrm{e}^2}
	\sum_{\boldsymbol{pq}} |v_q^{ \rm ei}|^2 {q^2 \over 3} f(E^\mathrm{e}_p)
        n_\mathrm{i}  \Gamma^{\rm V}(p) 
        {1\over (E^\mathrm{e}_p-E^\mathrm{e}_{p+q})^2} 
        \left[ {1 \over \hbar \omega + i 
        \eta - {\hbar^2 p q z \over m_\mathrm{e}} - 
        {\hbar^2 q^2 \over 2 m}}  + 
        {1 \over \hbar \omega + i  \eta + 
        {\hbar^2 p q z \over m_\mathrm{e}} + 
        {\hbar^2 q^2 \over 2 m_\mathrm{e}}}\right]\;, 
\end{equation}
and thereby
\begin{eqnarray}
        &&{\rm Re}\,\big\langle \,\,\dot{\!\!J}_{0}^{z} , 
        \,\,\dot{\!\!J}_{0}^{z} 
        \big\rangle_{\omega+i \eta}^{\rm V} =   
        { e^2 \hbar \over \beta \Omega_0 m_\mathrm{e}^2}
        {n_\mathrm{i}  \pi \over ( \hbar \omega )^2} 
	\sum_{\boldsymbol{pq}} |v_q^{ \rm ei}|^2 {q^2 \over 3} f(E^\mathrm{e}_p) 
        \Gamma^{\rm V}(p)\left[\delta (\hbar \omega - 
        {\hbar^2 p q z \over m_\mathrm{e}} -
        {\hbar^2 q^2 \over 2 m_\mathrm{e}}) + 
        \delta (\hbar \omega + {\hbar^2 p q z \over m_\mathrm{e}} +
        {\hbar^2 q^2 \over 2 m_\mathrm{e}})\right]\nonumber\\
        &&= 
        { e^6 \over \epsilon_0^2 \beta \hbar m_\mathrm{e} }
        {1  \over ( \hbar \omega)^2}   
        {n_\mathrm{i}
        n_\mathrm{e} \Lambda_\mathrm{e}^3 \over 24 \pi^3 \Omega_0} 
        \int_0^{\infty}\!  \mathrm{d}p\,p\, 
        \mathrm{e}^{-\beta \hbar^2 p^2/2m_\mathrm{e}} 
        \Gamma^{\rm V}(p) \int_0^{\infty}\! \mathrm{d}q\, {q^3 \over (q^2 + \kappa^2)^2} 
        \int\limits_{-1}^1 \mathrm{d} z
        \left[\delta ( z + {q \over 2p} - 
        {m_\mathrm{e} \omega \over \hbar p q} ) + 
        \delta ( z + {q \over 2p} + 
        {m_\mathrm{e} \omega \over \hbar p q} ) \right]~.
        \nonumber\\
\end{eqnarray}
The corresponding Green function $G_{\dot J\dot J}(\omega _\mu)$ evaluates to Eq.~(\ref{eqn:vert_green}).
\end{appendix}

\newpage
\end{document}